\newcommand*{\ATLASLATEXPATH}{./}
\documentclass[UKenglish,texlive=2011,cernpreprint,txfonts]{\ATLASLATEXPATH atlasdoc}
\pdfoutput=1 

\usepackage[biblatex=false,backend=bibtex]{\ATLASLATEXPATH atlaspackage}
\usepackage{\ATLASLATEXPATH atlascontribute}
\usepackage{\ATLASLATEXPATH atlasphysics}

\graphicspath{{logos/}{figures/}{./}{figs/}}

\usepackage{TOPQ_2013_02-defs}

\usepackage{cite}

\hypersetup{pdftitle={ATLAS paper},pdfauthor={The ATLAS Collaboration}}
\AtlasTitle{Measurement of the top quark mass in the \ttbarlj\ and
  \ttbarll\ channels using $\sqrt{s}=7$~\TeV\ ATLAS data}

\author{The ATLAS Collaboration}

\AtlasRefCode{TOPQ-2013-02}

\PreprintIdNumber{CERN-PH-EP-2015-050}

\AtlasJournal{\EPJC}

\AtlasAbstract{%
 The top quark mass was measured in the channels \ttbarlj\ and \ttbarll\
  (lepton=$e, \mu$) based on ATLAS data recorded in 2011.
  The data were taken at the LHC with a proton--proton centre-of-mass energy of
  $\rts=7$~\TeV\ and correspond to an integrated luminosity of \atlumo~\ifb.
  The \ttbarlj\ analysis uses a three-dimensional template technique which
  determines the top quark mass together with a global jet energy scale factor~(\JSF),
  and a relative $b$-to-light-jet energy scale factor~(\bJSF), where the terms \bjets\
  and light-jets refer to jets originating from $b$-quarks and $u, d, c,
  s$-quarks or gluons, respectively.
  The analysis of the \ttbarll\ channel exploits a one-dimensional template
  method using the \mlb\ observable, defined as the average invariant mass of
  the two lepton+\bjet\ pairs in each event.
  The top quark mass is measured to be $172.33\pm 0.75(\rm {stat + \JSF + \bJSF}) \pm 1.02(\rm
  {syst})$~\GeV, and $173.79 \pm 0.54({\rm stat}) \pm 1.30({\rm syst})$~\GeV\ in
  the \ttbarlj\ and \ttbarll\ channels, respectively.  The combination of the
  two results yields $\mt = 172.99 \pm 0.48({\rm stat}) \pm 0.78({\rm
  syst})$~\GeV, with a total uncertainty of~$0.91$~\GeV. 
}

\begin{document}

\maketitle

\tableofcontents

\newpage

\section{Introduction}
\label{sec:intro}
%
The mass of the top quark (\mt) is an important parameter of the Standard Model
(SM) of particle physics.
Precise measurements of \mt\ provide critical inputs to fits of global
electroweak parameters~\cite{LEPewkfits, GFitterNEW, PDG2014} that help assess
the internal consistency of the SM. In addition, the value of
\mt\ affects the stability of the SM Higgs potential, which has
cosmological implications~\cite{HiggsStab, HiggsStab1, HiggsStab2}.
 
Many measurements of \mt\ were performed by the CDF and D0 collaborations based
on Tevatron proton--antiproton collision data corresponding to integrated
luminosities of up to \tevlum~fb$^{-1}$.
A selection of these measurements was used in the recent Tevatron \mt\
combination resulting in $\mt = 174.34 \pm 0.37 \mbox{\thinspace (stat)} \pm
0.52 \mbox{\thinspace (syst)} \GeV = 174.34 \pm 0.64$~\GeV~\cite{TEV2014}.
Since 2010, measurements of \mt\ from the LHC by the ATLAS and CMS
collaborations have become available. They are based on proton--proton~($pp$)
collisions at a centre-of-mass energy of $\sqrt{s} = 7~\TeV$, recorded during
2010 and 2011 for integrated luminosities of up to
$4.9$~fb$^{-1}$~\cite{ATL-2012-Mass,CMSlj2011,CMSdil2011,CMSendpoint2011,CMSaj2011,ATLASaj2014}.
The corresponding LHC combination, based on $\sqrt{s} = 7~\TeV$ data and
including preliminary results, yields $\mt = 173.29 \pm 0.23
\mbox{\thinspace (stat)} \pm 0.92 \mbox{\thinspace (syst)} \GeV =
173.29 \pm 0.95$~\GeV~\cite{LHC2013}.
Using the same LHC input measurements and a selection of the \mt\ results from
the Tevatron experiments, the first Tevatron+LHC \mt\ combination results in
$\mt = 173.34 \pm 0.27 \mbox{\thinspace (stat)}
\pm 0.71 \mbox{\thinspace (syst)}$ GeV, with a total uncertainty of
$0.76$~\GeV~\cite{WA2014}.
Recently, improved individual measurements with a total uncertainty compatible
with that achieved in the Tevatron+LHC \mt\ combination have become available;
the most precise single measurement is obtained by the D0 Collaboration
using \ttbarlj\ events and yields $\mt = 174.98\pm 0.76~\GeV$~\cite{D0lj2014}.

This article presents a measurement of \mt\ using events with one or two isolated
charged leptons~(electrons or muons) in the final state (the \ttbarlj\ and
\ttbarll\ decay channels), in \atlumo~\ifb\ of $pp$ collision data
collected by the ATLAS detector at a centre-of-mass energy of $\rts=7$~\TeV\
during 2011. It supersedes Ref.~\cite{ATL-2012-Mass}, where, using a
two-dimensional fit to reconstructed observables in the \ttbarlj\ channel, 
\mt\ was determined together with
a global jet energy scale factor.
The use of this scale factor allows the uncertainty on \mt\ stemming from 
imperfect knowledge of the jet energy scale~(JES) to be considerably reduced, 
albeit at the cost of an additional statistical uncertainty component.
The single largest systematic uncertainty on \mt\ in Ref.~\cite{ATL-2012-Mass}
was due to the relative $b$-to-light-jet energy scale (bJES) uncertainty, where
the terms \bjets\ and light-jets refer to jets originating from \bquarks\ and
$u, d, c, s$-quarks or gluons, respectively.
To reduce this uncertainty in the present analysis, a three-dimensional template
fit is used for the first time in the \ttbarlj\ channel, again replacing the
corresponding uncertainty by a statistical uncertainty and a reduced systematic
uncertainty. This concept will be even more advantageous with increasing data
luminosity.
In addition, for the combination of the measurements of \mt\ in the two decay
channels an in-depth investigation of the correlation of the two estimators for
all components of the sources of systematic uncertainty is made. This leads to a
much smaller total correlation of the two measurements than what is typically
assigned, such that their combination yields a very significant improvement in
the total uncertainty on \mt.
To retain this low correlation, the jet energy scale factors measured in the
\ttbarlj\ channel have not been propagated to the \ttbarll\ channel.

In the \ttbarlj\ channel, one $W$~boson from the top or antitop quark decays
directly or via an intermediate $\tau$ decay into an electron or muon and at
least one neutrino, while the other $W$ boson decays into a quark--antiquark
pair. The \ttbar\ decay channels with electrons and muons are combined and
referred to as the lepton+jets (or as a shorthand \ljets) final state.
The \ttbarll\ channel corresponds to the case where both $W$ bosons from the top
and antitop quarks decay leptonically, directly or via an intermediate $\tau$
decay, into an electron or muon and at least one neutrino. The \ttbar\ decay
channels $ee, \emu, \mu\mu$ are combined and referred to as the \dil\ final
state.  For both the \ljets\ and \dil\ final states, the measurements are based
on the template method~\cite{CDFlj2006}.
In this technique, Monte Carlo (MC) simulated distributions are constructed for
a chosen quantity sensitive to the physics parameter under study, using a number
of discrete values of that parameter.
These templates are fitted to analytical functions that interpolate between
different input values of the physics parameter, fixing all other parameters of
the functions.
In the final step a likelihood fit to the observed distribution in data is used
to obtain the value for the physics parameter that best describes the data.
In this procedure the top quark mass determined from data corresponds to the
mass definition used in the MC simulation. It is expected that the difference between
this mass definition and the pole mass is of order
1~GeV~\cite{BUC-1101,Hoang2008,Hoang2014,MITP2014}.

In the \ljets\ channel, events are reconstructed using a kinematic fit that
assumes a \ttbar\ topology. A three-dimensional template method is used,
where \mt\ is determined simultaneously with a light-jet energy scale factor
(\JSF), exploiting the information from the hadronic $W$ decays, and a separate
$b$-to-light-jet energy scale factor (\bJSF).
The \JSF\ and \bJSF\ account for residual differences of data and simulation in
the light-jet and in the relative $b$-to-light-jet energy scale, respectively,
thereby mitigating the corresponding systematic uncertainties on \mt.
The analysis in the \dil\ channel is based on a one-dimensional template method,
where the templates are constructed for the \mlb\ observable, defined as the
per-event average invariant mass of the two lepton$-b$-jet systems from the
decay of the top quarks.
Due to the underconstrained kinematics associated with the \dil\ final state, no
in situ constraint of the jet energy scales is performed.

This article is organised as follows: after a short description of the ATLAS
detector in Sect.~\ref{sec:atlas}, the data and MC simulation samples are
discussed in Sect.~\ref{sec:dataMC}. Details of the event selection and
reconstruction are given in Sect.~\ref{sec:evtselrec}.
The template fits are explained in Sect.~\ref{sec:templates}.
The measurement of \mt\ in the two final states is given in
Sect.~\ref{sec:result}, and the evaluation of the associated systematic
uncertainties are discussed in Sect.~\ref{sec:syst}. The results of the
combination of the \mt\ measurements from the individual analyses are reported
in Sect.~\ref{sec:resultcomb}.
Finally, the summary and conclusions are given in Sect.~\ref{sec:summary}.


\section{The ATLAS detector}
\newcommand{\AtlasCoordFootnote}{%
ATLAS uses a right-handed coordinate system with its origin at the
nominal interaction point (IP) in the centre of the detector and the
$z$-axis along the beam pipe.  The $x$-axis points from the IP to the
centre of the LHC ring, and the $y$-axis points upwards.  Cylindrical
coordinates $(r,\phi)$ are used in the transverse plane, $\phi$ being
the azimuthal angle around the beam pipe.  The pseudorapidity is
defined in terms of the polar angle $\theta$ as $\eta = -\ln
\tan(\theta/2)$.  Angular distances are defined as $\Delta R \equiv
\sqrt{(\Delta\eta)^{2} + (\Delta\phi)^{2}}$.}

\label{sec:atlas}

The ATLAS detector~\cite{ATL-2008-001} covers nearly the entire solid angle
around the collision point.\footnote{\AtlasCoordFootnote} It consists of an
inner tracking detector surrounded by a thin superconducting solenoid,
electromagnetic and hadronic calorimeters, and a muon spectrometer incorporating
three large superconducting toroid magnets.  The inner-detector system (ID) is
immersed in a \SI{2}{\tesla} axial magnetic field and provides charged-particle
tracking in the range $|\eta| < 2.5$.  The high-granularity silicon pixel
detector covers the interaction region and typically provides three measurements
per track, the first energy deposit being normally in the innermost layer.  It
is followed by the silicon microstrip tracker designed to provide four
two-dimensional measurement points per track.  These silicon detectors are
complemented by the transition radiation tracker, which enables radially
extended track reconstruction up to $|\eta| = 2.0$.  The transition radiation
tracker also provides electron identification information based on the fraction
of energy deposits (typically 30 hits in total) above an energy threshold
corresponding to transition radiation.  The calorimeter system covers the
pseudorapidity range $|\eta| < 4.9$.  Within the region $|\eta|< 3.2$,
electromagnetic calorimetry is provided by barrel and endcap high-granularity
lead/liquid-argon (LAr) electromagnetic calorimeters, with an additional thin
LAr presampler covering $|\eta| < 1.8$, to correct for energy loss in material
upstream of the calorimeters.  Hadronic calorimetry is provided by the
steel/scintillator-tile calorimeter, segmented into three barrel structures
within $|\eta| < 1.7$, and two copper/LAr hadronic endcap calorimeters.  The
solid angle coverage is completed with forward copper/LAr and tungsten/LAr
calorimeter modules optimised for electromagnetic and hadronic measurements
respectively.  The muon spectrometer (MS) comprises separate trigger and
high-precision tracking chambers measuring the deflection of muons in the magnetic
field generated by the toroids.  The precision chamber
system covers the region $|\eta| < 2.7$ with three layers of monitored drift
tubes, complemented by cathode strip chambers in the forward region.  The muon
trigger system covers the range $|\eta| < 2.4$ with resistive plate chambers in
the barrel, and thin gap chambers in the endcap regions.  A three-level trigger
system is used to select interesting events~\cite{atlas-trigger-2010}.  The
Level-1 trigger is implemented in hardware and uses a subset of detector
information to reduce the event rate to at most \SI{75}{\kHz}.  This is followed
by two software-based trigger levels which together reduce the event rate to
about 300~Hz.

\section{Data and Monte Carlo samples}
\label{sec:dataMC}
%
For the measurements described in this document, data from LHC $pp$ collisions
at $\rts=7$~\TeV\ are used.
They correspond to an integrated luminosity of \atlumo~\ifb\ with an uncertainty
of \atlumounc~\cite{Aad:2013ucp}, and were recorded during 2011 during stable
beam conditions and with all relevant ATLAS sub-detector systems operational.

MC simulations are used to model \ttbar\ and single top quark processes as well
as some of the background contributions.  Top quark pair and single top quark
production (in the $s$- and $Wt$-channels) are simulated using the
next-to-leading-order (NLO) MC program \Powheg-hvq (patch4)~\cite{FRI-0701} with
the NLO CT10~\cite{LAI-1001} parton distribution functions (PDFs).
Parton showering, hadronisation and the underlying event are modelled using
the \Pythia\ (v6.425)~\cite{SJO-0601} program with the Perugia 2011C (P2011C)
MC parameter set~(tune)~\cite{Skands} and the corresponding CTEQ6L1 PDFs~\cite{cteq6l}.
The {\sc AcerMC} (v3.8) generator~\cite{SAMPLES-ACER} interfaced with
\Pythia (v6.425) is used for the simulation of the single top quark
$t$-channel process. The \Acermc\ and \Pythia\ programs are used with the
CTEQ6L1 PDFs and the corresponding P2011C tune.

For the construction of signal templates, the \ttbar\ and single top quark
production samples are generated for different assumed values of
\mt, namely $167.5, 170, 172.5, 175, 177.5~\GeV$.
The \ttbar\ MC samples are normalised to the predicted
\ttbar\ cross section for each \mt\ value.  The $t\bar{t}$ cross
section for $pp$ collisions at $\sqrt{s} = 7 \tev$ is $\sigma_{t\bar{t}}=
177^{+10}_{-11}$~pb for $\mt=172.5$~\gev.
It was calculated at next-to-next-to-leading-order (NNLO) in QCD including
resummation of next-to-next-to-leading-logarithmic (NNLL) soft gluon terms with
Top++2.0~\cite{CAC-1101,PRL-109-132001,JHEP-1212,JHEP-1301,TopPP20,CZA-1101}.
The PDF+\al\ uncertainties on the cross section were calculated using the
PDF4LHC prescription~\cite{PDF4LHC} with the MSTW2008 $68\%$
CL~NNLO~\cite{MAR-0901,MAR-0902}, CT10~NNLO~\cite{LAI-1001,CT10NNLO} and
NNPDF2.3~5f~FFN~\cite{Ball:2012cx} PDFs, and added in quadrature to the
factorisation and renormalisation scale uncertainty.  The NNLO+NNLL value, as implemented in
Hathor~1.5~\cite{ALI-1101}, is about $3\%$ larger than the plain NNLO
prediction.
The single top quark production cross sections are normalised to the approximate
NNLO prediction values. For example, for $\mt=172.5$~\gev, these are
$64.6^{+2.7}_{-2.0}$ pb \cite{Kidonakis:2011wy}, $4.6\pm 0.2$ pb
\cite{Kidonakis:2010tc} and $15.7\pm 1.1$ pb \cite{Kidonakis:2010ux}
for the $t$-, $s$- and $Wt$- production channels respectively.

The production of \Wboson\ or \Zboson\ bosons in association with jets is
simulated using the \Alpgen\ (v2.13) generator~\cite{MAN-0301} interfaced to
the \Herwig\ (v6.520)~\cite{Marchesini, COR-0001} and
\Jimmy\ (v4.31)~\cite{SAMPLES-JIMMY} packages. The CTEQ6L1 PDFs and
the corresponding AUET2 tune~\cite{ATL-PHYS-PUB-2011-008} are used for the
matrix element and parton shower settings. The $W$+jets events containing
heavy-flavour quarks ($Wbb$+jets, $Wcc$+jets, and $Wc$+jets) are generated
separately using leading-order matrix elements with massive $b$- and
\cquarks. An overlap-removal procedure is used to avoid double counting of
heavy-flavour quarks between the matrix element and the parton shower evolution.
Diboson production processes ($WW$, $WZ$ and $ZZ$) are produced using
the \Herwig\ generator with the AUET2 tune.

Multiple $pp$ interactions generated with \Pythia (v6.425) using the AMBT2B
tune~\cite{ATL-PHYS-PUB-2011-009} are added to all MC samples.  These simulated
events are re-weighted such that the distribution of the number of interactions
per bunch crossing (pile-up) in the simulated samples matches that in the data.
The average number of interactions per bunch crossing for the data set
considered is 8.7.
The samples are processed through a simulation of the ATLAS
detector~\cite{WT-ATLAS-SIMULATION-PAPER} based on GEANT4~\cite{AGO-0301} and
through the same reconstruction software as the data.

\section{Event selection and reconstruction}
\label{sec:evtselrec}
\subsection{Object selection}
In this analysis \ttbar\ events with one or two isolated charged leptons in the
final states are selected.
The event selection for both final states is based on the following
reconstructed objects in the detector: electron and muon candidates, jets and
missing transverse momentum (\met).

An electron candidate is defined as an energy deposit in the electromagnetic
calorimeter with an associated well-reconstructed
track~\cite{CERN-PH-EP-2014-040}.
Electron candidates are required to have transverse energy $\ET>25$~\GeV\ and
$\absetaclus < 2.47$, where \etaclus\ is the pseudorapidity of the
electromagnetic cluster associated with the electron.
Candidates in the transition region between the barrel and endcap calorimeter
($1.37<\absetaclus<1.52$) are excluded.
Muon candidates are reconstructed from track segments in different layers of the
MS~\cite{CERN-PH-EP-2014-151}. These segments are combined starting from the
outermost layer, with a procedure that takes effects of detector material into
account, and matched with tracks found in the ID. The final candidates are
refitted using the complete track information, and are required to satisfy
$\pt>20$~\GeV\ and $\vert\eta\vert<2.5$.
Isolation criteria, which restrict the amount of energy deposited near the
lepton candidates, are applied to both the electrons and muons to reduce the
backgrounds from heavy-flavour decays inside jets or photon conversions,
and the background from hadrons mimicking lepton signatures, in the following
referred to as non-prompt and fake-lepton background (NP/fake-lepton
background).
For electrons, the energy not associated with the electron cluster and contained
in a cone of $\Delta R = 0.2$ around the electron must not exceed an
$\eta$-dependent threshold ranging from 1.25 to 3.7~\GeV.  Similarly, the total
transverse momentum of the tracks contained in a cone of $\Delta R=0.3$ must not
exceed a threshold ranging from 1.00 to 1.35~\GeV, depending on the electron
candidate \pt\ and $\eta$.
For muons, the sum of track transverse momenta in a cone of $\Delta R=0.3$
around the muon is required to be less than 2.5~\GeV, and the total energy
deposited in a cone of $\Delta R=0.2$ around the muon is required to be less
than \isoenmu~\GeV.  The longitudinal impact parameter of each charged lepton
along the beam axis is required to be within 2~mm of the reconstructed primary
vertex, defined as the vertex with the highest $\sum_{\rm trk} p_{\rm T,trk}^2$,
among all candidates with at least five associated tracks with $p_{\rm T,trk} >
0.4~\GeV$.

Jets are reconstructed with the anti-\kt algorithm~\cite{CAC-0801}
using a radius parameter of $R=0.4$, starting from energy clusters
of adjacent calorimeter cells called topological
clusters~\cite{Lampl:1099735}.
These jets are calibrated first by correcting the jet energy using the scale
established for electromagnetic objects (EM scale).  They are further corrected
to the hadronic energy scale using calibration factors that depend on the jet
energy and $\eta$, obtained from simulation.  Finally, a residual in situ
calibration derived from both data and MC simulation is
applied~\cite{CERN-PH-EP-2013-222}.
Jet quality criteria are applied to identify and reject jets
reconstructed from energy deposits in the calorimeters originating
from particles not emerging from the bunch crossing under
study~\cite{ATLAS-CONF-2012-20}.
To suppress the contribution from low-\pt\ jets originating from
pile-up interactions, tracks associated with the jet and emerging from
the primary vertex are required to account for at least 75\% of the
scalar sum of the \pt\ of all tracks associated with the jet. Jets
with no associated tracks are also accepted.

Muons reconstructed within a $\Delta R=0.4$ cone around a jet satisfying
$\pt>25$~\GeV\ are removed to reduce the contamination caused by muons from
hadron decays within jets.
Subsequently, jets within a $\Delta R=0.2$ cone around an electron candidate are
removed to avoid double counting, which can occur because electron clusters are
usually also reconstructed as jets.
After this jet overlap removal, electrons are rejected if their distance to the
closest jet is smaller than $\Delta R=0.4$.

The reconstruction of \met\ is based on the vector sum of calorimeter energy
deposits projected onto the transverse plane.
The \met\ is reconstructed from topological clusters, calibrated at the EM scale
and corrected according to the energy scale of the corresponding identified physics
objects. Contributions from muons are included by using their momentum as
measured by the inner detector and muon spectrometer~\cite{ATLAS-MET-NEW}.

The reconstruction of top quark pair events is facilitated by the ability to tag
jets originating from \bquarks.
For this purpose the neural-network-based MV1 algorithm is
applied~\cite{ATLAS-CONF-2012-040,ATLAS-CONF-2012-043}.
In the following, irrespective of their origin, jets tagged by this algorithm
are called \btagged\ jets, whereas those not tagged are called untagged jets.
Similarly, whether they are tagged or not, jets originating from \bquarks\ and
from $u, d, c, s$-quarks or gluons are called \bjets\ and light-jets,
respectively.
The MV1 algorithm relies on track impact parameters and the properties of
reconstructed secondary vertices such as the decay length significance.
The chosen working point corresponds to a \btag\ efficiency of \btageff\ for
\bjets\ in simulated \ttbar\ events and a light-jet ($c$-quark jet) rejection
factor of \btagrej (4).
To match the \btag\ performance in the data, \pt- and $\eta$-dependent scale
factors are applied to MC jets depending on their original flavour. The scale
factors are obtained from dijet~\cite{ATLAS-CONF-2012-043} and
\ttbarll\ events. The \ttbar-based calibration is obtained using the methodology
described in Ref.~\cite{ATLAS-CONF-2014-004}, applied to the 7~\TeV\ data. The
scale factors are calculated per jet and finally multiplied to obtain an event
weight for any reconstructed distribution.
%
\subsection{Event selection}
The \ttbarlj\ signal is characterised by an isolated charged lepton with
relatively high \pt, \met\ arising from the neutrino from the leptonic
\Wboson\ boson decay, two \bjets\ and two light-jets from the hadronic
\Wboson\ boson decay.  The main contributions to the background stem from
\Wj\ production and from the NP/fake-lepton background.
The normalisation of the \Wj\ background is estimated from data, based on the
charge-asymmetry method~\cite{CERN-PH-EP-2012-015}, and the shape is obtained
from simulation.
For the NP/fake-lepton background, both the shape of the distributions and the normalisation are
estimated from data by weighting each selected event by the probability of
containing a NP/fake lepton. This contribution in both the electron and
the muon channel is estimated using a data-driven matrix method based on
selecting two categories of events, using loose and tight lepton selection
requirements~\cite{Aad:2010ey}.
The contributions from single top quark, $Z$+jets, and diboson production are
taken from simulation, normalised to the best available theoretical cross
sections.

The \ttbarll\ events are characterised by the presence of two isolated and
oppositely charged leptons with relatively high \pt, \met\ arising from the
neutrinos from the leptonic \Wboson\ boson decays, and two \bjets. Background
processes with two charged leptons from \Wboson- or \Zboson\ decays in the final
state, which are similar to the \ttbarll\ events, are dominated by single top
quark production in the $Wt$-channel.
Additional contributions come from $Z$+jets processes and diboson production
with additional jets.  In the analysis, these contributions are estimated
directly from the MC simulation normalised to the relevant cross sections.
Events may also be wrongly reconstructed as \ttbarll\ events due to the presence
of NP/fake leptons together with \btagged\ jets and \MET.  As for the
\ttbarlj\ channel, the NP/fake-lepton background is estimated using a
data-driven matrix method~\cite{Aad:2010ey}.

The selection of \ttbar\ event candidates consists of a series of requirements
on the general event quality and the reconstructed objects designed to select
events consistent with the above signal topologies.  To suppress non-collision
background, events are required to have at least one good primary vertex. 
It is required that the appropriate single-electron or single-muon trigger has
fired; the trigger thresholds are 20 or 22~\GeV\ (depending on the data-taking
period) for the electrons and 18~\GeV\ for muons.
Candidate events in the \ljets\ final state are required to have exactly one
reconstructed charged lepton with $\ET> 25$~\GeV\ for electrons, and $\pt >
20$~\GeV\ for muons, matching the corresponding trigger object. Exactly two
oppositely charged leptons, with at least one matching a trigger object, are
required in the \dil\ final state.
In the \mjets\ channel, $\met>20$~\GeV\ and $\met+\mWt>60$~\GeV\ are
required.\footnote{Here \mWt\ is the \Wboson\ boson transverse mass, defined as
  $\sqrt{2\,p_\mathrm{T,\ell}\,p_\mathrm{T,\nu}\left[1-\cos(\phi_{\ell}-\phi_{\nu})\right]}$,
  where the measured \met\ vector provides the neutrino ($\nu$) information.}
In the \ejets\ channel more stringent selections on \met\ and \mWt\ ($\met >
30$~\GeV\ and $\mWt>30$~\GeV) are imposed due to the higher level of
NP/fake-lepton background.  For the $ee$ and $\mu\mu$ channels, in the
\dil\ final state, $\MET>60~\GeV$ is required.  In addition, the invariant mass
of the same-flavour charged-lepton pair, $m_{\ell\ell}$~$(\ell\ell= ee, \mu\mu)$, is required to exceed
15~\GeV, to reduce background from low-mass resonances decaying into charged
lepton--antilepton pairs and Drell--Yan production. Similarly, to reduce the
$Z$+jets background, values of $m_{\ell\ell}$ compatible with the
\Zboson\ boson mass are vetoed by requiring $|m_{\ell\ell} - 91~\GeV| >
10~\GeV$.  In the \emu channel $\Ht>130~\GeV$ is required, where \Ht\ is the
scalar sum of the $\pt$ of the two selected charged leptons and the jets.
Finally, the event is required to have at least four jets (or at least two jets
for the \ttbarll\ channel) with $\pt>25$~\GeV\ and $\vert\eta\vert<2.5$.
At least one of these jets must be \btagged\ for the \ttbarlj\ analysis. In the
\dil\ final state, events are accepted if they contain exactly one or two
\btagged\ jets.

These requirements select 61786 and 6661 data events in the \ttbarlj\ and
\ttbarll\ channels, with expected background fractions of 22\% and 2\%,
respectively.  Due to their inherent \mt\ sensitivity, here and in the
following, the single top quark processes are accounted for as signal in both
analyses, and not included in the quoted background fractions.
%
\subsection{Event reconstruction}
After the event selection described in the previous section, the events are
further reconstructed according to the decay topology of interest, and are
subject to additional requirements.
%
\subsubsection{Kinematic reconstruction of the lepton+jets final state}
A kinematic likelihood fit~\cite{ATL-2012-Mass,KLFitterNIM} is used to fully
reconstruct the \ttbarlj\ kinematics. The algorithm relates the measured
kinematics of the reconstructed objects to the leading-order representation of
the \ttbar\ system decay. The event likelihood is constructed as the product of
Breit--Wigner (BW) distributions and transfer functions (TF). The $W$ boson BW
line-shape functions use the world combined values of the $W$ boson mass and
decay width from Ref.~\cite{PDG2014}.
A common mass parameter, \mtr, is used for the BW distributions describing the
leptonically and hadronically decaying top quarks, and this is fitted
event-by-event.  The top quark width varies with \mtr\ and it is calculated
according to the SM prediction~\cite{PDG2014}.  The TF are derived from the
\Powheg+\Pythia\ \ttbar\ signal MC simulation sample at an input mass of
$\mt=172.5$~\GeV.
They represent the experimental resolutions in terms of the probability that the
observed energy at reconstruction level is produced by a given parton-level
object for the leading-order decay topology.

The input objects to the likelihood are: the reconstructed charged lepton, the
missing transverse momentum and four jets. For the sample with one \btagged\ jet
these are the \btagged\ jet and the three untagged jets with the highest
\pt. For the sample with at least two \btagged\ jets these are the two
highest-\pt\ \btagged\ jets, and the two highest-\pt\ remaining jets. The $x$-
and $y$-components of the missing transverse momentum are used as starting
values for the neutrino transverse momentum components, with its longitudinal
component ($p_{\nu,z}$) as a free parameter in the kinematic likelihood fit. Its
starting value is computed from the $W\to\ell\nu$ mass constraint. If there are
no real solutions for $p_{\nu,z}$ a starting value of zero is used.  If there
are two real solutions, the one giving the largest likelihood value is taken.

Maximising the event-by-event likelihood as a function of \mtr\ establishes the
best assignment of reconstructed jets to partons from the \ttbarlj\ decay. The
maximisation is performed by testing all possible permutations, assigning jets
to partons. The likelihood is extended by including the probability for a jet to
be \btagged, given the parton from the top quark decay it is associated with, to
construct an event probability.  The $b$-tagging efficiencies and rejection
factors are used to favour permutations for which a \btagged\ jet is assigned to
a $b$-quark and penalise those where a \btagged\ jet is assigned to a light
quark. The permutation of jets with the highest likelihood value is retained.

The value of \mtr\ obtained from the kinematic likelihood fit is used as the
observable primarily sensitive to the underlying \mt.  The invariant mass of the
hadronically decaying \Wboson\ boson (\mWr) is calculated from the assigned jets
of the chosen permutation.  Finally, an observable called \rlbr, designed to be
sensitive to the relative $b$-to-light-jet energy scale, is computed in the
following way.  For events with only one \btagged\ jet, \rlbr\ is defined as the
ratio of the transverse momentum of the \btagged\ jet to the average transverse
momentum of the two jets of the hadronic \Wboson\ boson decay.  For events with
two or more \btagged\ jets, \rlbr\ is defined as the scalar sum of the
transverse momenta of the \btagged\ jets assigned to the leptonically and
hadronically decaying top quarks divided by the scalar sum of the transverse
momenta of the two jets associated with the hadronic \Wboson\ boson decay.
The values of \mWr\ and \rlbr\ are computed from the jet four-vectors as given
by the jet reconstruction to keep the maximum sensitivity to changes of the jet
energy scale for light-jets and \bjets.

In view of the template parameterisation described in Sect.~\ref{sec:templates}
additional selection criteria are applied.  Events in which a \btagged\ jet is
assigned to the \Wboson\ decay by the likelihood fit are discarded. This is
needed to prevent mixing effects between the information provided by the
\mWr\ and \rlbr\ distributions.
The measured \mtr\ is required to be in the range 125~\GeV\ to 225~\GeV\ for
events with one \btagged\ jet, and in the range 130~\GeV\ to 220~\GeV\ for
events with at least two \btagged\ jets.
In addition, \mWr\ is required to be in the range 55~\GeV\ to
110~\GeV\ and finally, \rlbr\ is required to be in the range 0.3 to
3.0. 
The fraction of data events which pass these requirements is 35\%.
Although removing a large fraction of data, these requirements remove events in
the tails of the three distributions, which are typically poorly reconstructed
with small likelihood values and do not contain significant information on \mt.
In addition, the templates then have simpler shapes which are easier 
to model analytically with fewer parameters.
%
\subsubsection{Reconstruction of the \dil\ final state}
In the \ttbarll\ channel the kinematics are under-constrained due to the
presence of at least two undetected neutrinos. Consequently, instead of
attempting a full reconstruction, the \mt-sensitive observable \mlb\ is defined
based on the invariant mass of the two charged-lepton+\bjet\ pairs.

The preselected events contain two charged leptons, at least two jets, of which
either exactly one or exactly two are \btagged.
For events with exactly two \btagged\ jets the charged-lepton+\btagged\ jet
pairs can be built directly.
In the case of events with only one \btagged\ jet the missing second \bjet\ is
identified with the untagged jet carrying the highest MV1 weight.
For both classes of events, when using the two selected jets and the two charged
leptons, there are two possible assignments for the jet-lepton pairs, each
leading to two values for the corresponding pair invariant masses. The
assignment resulting in the lowest average mass is retained, and this mass is
taken as the \mlbr\ estimator of the event.  The measured \mlbr\ is required to
be in the range 30~\GeV\ to 170~\GeV. This extra selection retains 97\% of the
data candidate events.
%
\subsubsection{Event yields}
\label{sec:evtyields}
The numbers of events observed and expected after the above selections are
reported in Table~\ref{tab:LJDLcutflow} for the \ljets\ and \dil\ final states.
The observed numbers of events are well described by the sum of the signal and
background estimates within uncertainties. The latter are estimated as the sum
in quadrature of the statistical uncertainty, the uncertainty on the
\btag\ efficiencies, a \atlumounc\ uncertainty on the integrated
luminosity~\cite{Aad:2013ucp}, the uncertainties on the \ttbar\ and single top
quark theoretical cross sections, a {\ensuremath{30\%}\xspace} uncertainty on
the \Wj\ and \Zj\ normalisation, and finally a {\ensuremath{50\%}\xspace}
uncertainty on the NP/fake-lepton background normalisation.
The distribution of several kinematic variables in the data were inspected and
found to be well described by the signal-plus-background prediction, within
uncertainties. As examples, Fig.~\ref{fig:rec}(left) shows the distribution of
the untagged and \btagged\ jets \pt\ observed in the \ljets\ final
state. Similarly, the \pt\ distributions for the charged leptons and
\btagged\ jets in the \dil\ final state are shown on the right of
Fig.~\ref{fig:rec}. In all cases the data are compared with the MC predictions,
assuming an input top quark mass of 172.5~\GeV.
%
\begin{table}[tbp!]
\begin{center}
\small
\begin{tabular}{|l|rr|rr|rr|}
\hline
\multicolumn{7}{|c|}{\ljets\ final state} \\\hline
Process             & \multicolumn{2}{c|}{One \btagged\ jet} 
 & \multicolumn{2}{c|}{At least two \btagged\ jets} 
& \multicolumn{2}{c|}{Sum} \\  
\hline
$t\bar t$ signal    &  9890 $\pm$ &  630 & 8210 $\pm$ &  560 &  18100 $\pm$ & 1100 \\
Single top quark (signal) &   756 $\pm$ &   41 &  296 $\pm$ &   19 &   1052 $\pm$ &   57 \\\hline
$W$+jets (data)     &  2250 $\pm$ &  680 &  153 $\pm$ &   49 &   2400 $\pm$ &  730 \\
$Z$+jets            &   284 $\pm$ &   87 & 18.5 $\pm$ &  6.1 &    303 $\pm$ &   93 \\
$WW/WZ/ZZ$          &  43.5 $\pm$ &  2.3 & 4.65 $\pm$ & 0.48 &   48.2 $\pm$ &  2.6 \\
NP/fake leptons (data) &   700 $\pm$ &  350 &   80 $\pm$ &   41 &    780 $\pm$ &  390 \\
Signal+background   & 13920 $\pm$ & 1000 & 8760 $\pm$ &  560 &  22700 $\pm$ & 1400 \\\hline
Data                & \multicolumn{2}{c|}{12979} 
                    & \multicolumn{2}{c|}{8784} 
                    & \multicolumn{2}{c|}{21763} \\\hline
Exp. Bkg. frac.     & 0.25 $\pm$ & 0.02  & 0.03 $\pm$ & 0.00 &  0.16 $\pm$ & 0.01 \\
Data/MC             & 0.93 $\pm$ & 0.07  & 1.00 $\pm$ & 0.07 &  0.96 $\pm$ & 0.06 \\\hline\hline

\multicolumn{7}{|c|}{Dilepton final state} \\\hline
Process             & \multicolumn{2}{c|}{One \btagged\ jet} 
 & \multicolumn{2}{c|}{Two \btagged\ jets} 
& \multicolumn{2}{c|}{Sum} \\  
\hline
$t\bar t$ signal    &  2840 $\pm$ &   180  & 2950  $\pm$ &  210 &  5790 $\pm$ &   360\\
Single top quark (signal) &   181 $\pm$ &    10  & 82.5  $\pm$ &  5.7 &   264 $\pm$ &    15\\\hline
$Z$+jets            &    34 $\pm$ &    11  &  4.1  $\pm$ &  1.5 &    38 $\pm$ &    12\\
$WW/WZ/ZZ$          &  7.01 $\pm$ &  0.63  & 0.61  $\pm$ & 0.15 &  7.62 $\pm$ &  0.67\\
NP/fake leptons (data) &    52 $\pm$ &    28  &  2.6  $\pm$ &  8.4 &    55 $\pm$ &    30\\
Signal+background   &  3110 $\pm$ &   180  & 3040  $\pm$ &  210 &  6150 $\pm$ &   360\\\hline
Data                & \multicolumn{2}{c|}{3227} 
                    & \multicolumn{2}{c|}{3249} 
                    & \multicolumn{2}{c|}{6476} \\\hline
Exp. Bkg. frac.     & 0.03 $\pm$ & 0.00 & 0.00 $\pm$ & 0.00 & 0.02 $\pm$ & 0.00 \\
Data/MC             & 1.04 $\pm$ & 0.06 & 1.07 $\pm$ & 0.07 & 1.05 $\pm$ & 0.06 \\\hline
\end{tabular}
\end{center}
\caption{The observed numbers of events, according to the \btagged\ jet
  multiplicity, in the \ljets\ and \dil\ final states in \atlumo~\ifb\ of
  $\sqrt{s} = 7$~\TeV\ data. In addition, the expected numbers of signal and
  background events corresponding to the integrated luminosity of the data are
  given. The predictions are quoted using two significant digits for their
  uncertainty.
  The MC estimates assume SM cross sections. The \Wj\ and NP/fake-lepton
  background contributions are estimated from data. The uncertainties for the
  estimates include the components detailed in Sect.~\ref{sec:evtyields}.
  Values smaller than $0.005$ are listed as $0.00$.
  \label{tab:LJDLcutflow}}
\end{table}
%
\begin{figure*}[tbp!]
\centering
\subfloat[\ttbarlj: untagged jets from \Wboson\ boson decays]{
  \includegraphics[width=0.49\textwidth]{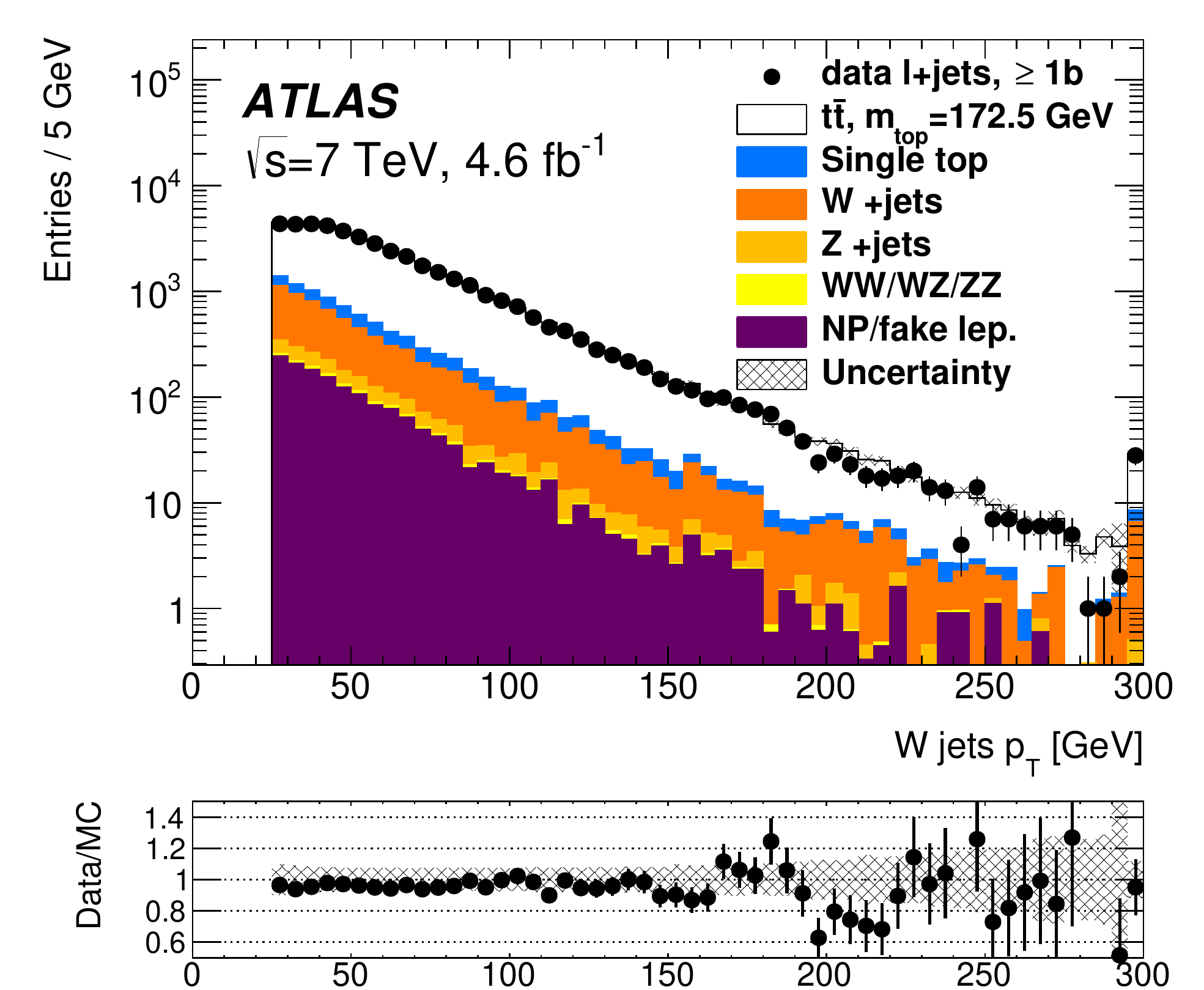}}
\subfloat[\ttbarll: charged leptons]{
  \includegraphics[width=0.49\textwidth]{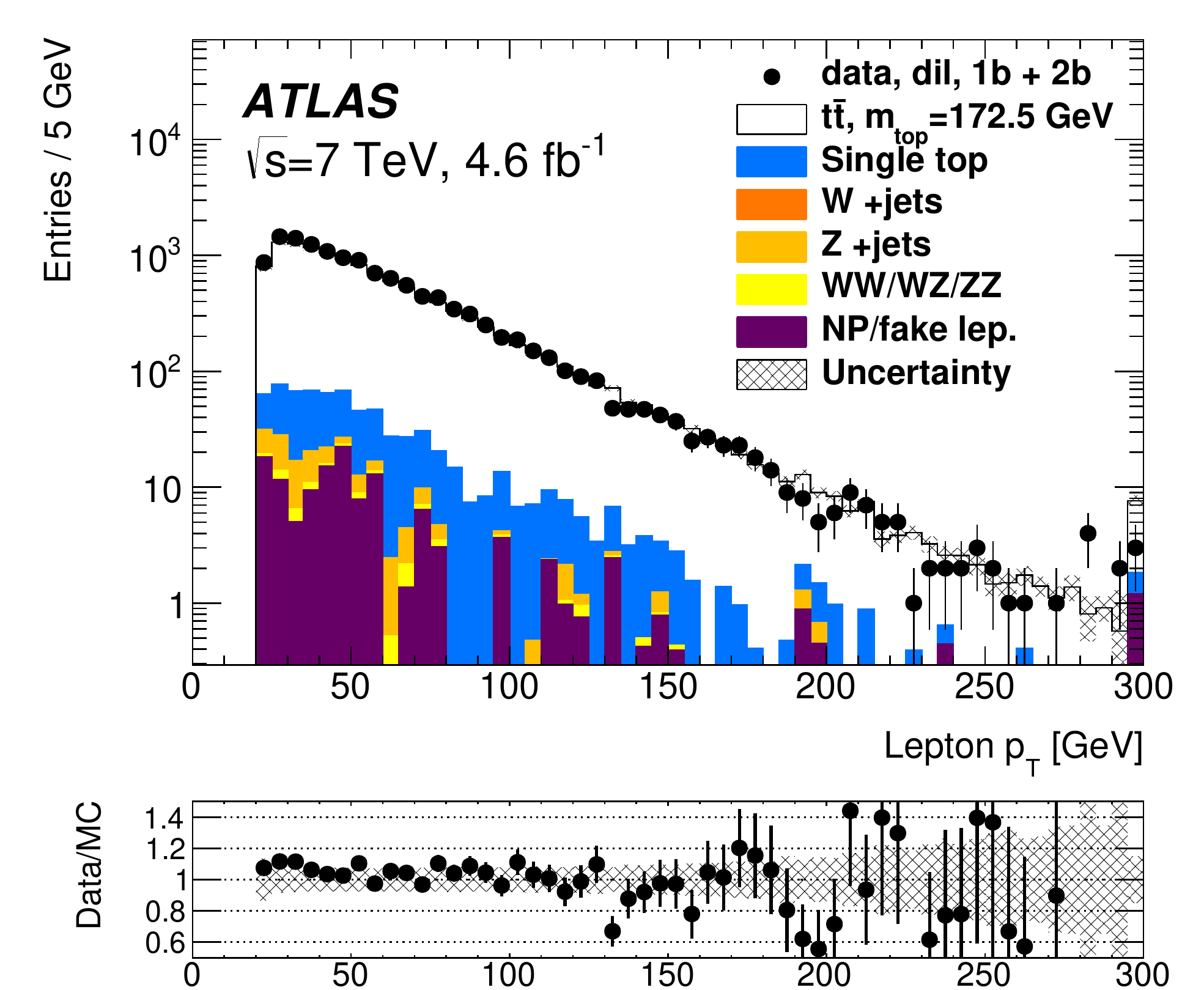}} \hfill
\subfloat[\ttbarlj: \btagged\ jets]{
  \includegraphics[width=0.49\textwidth]{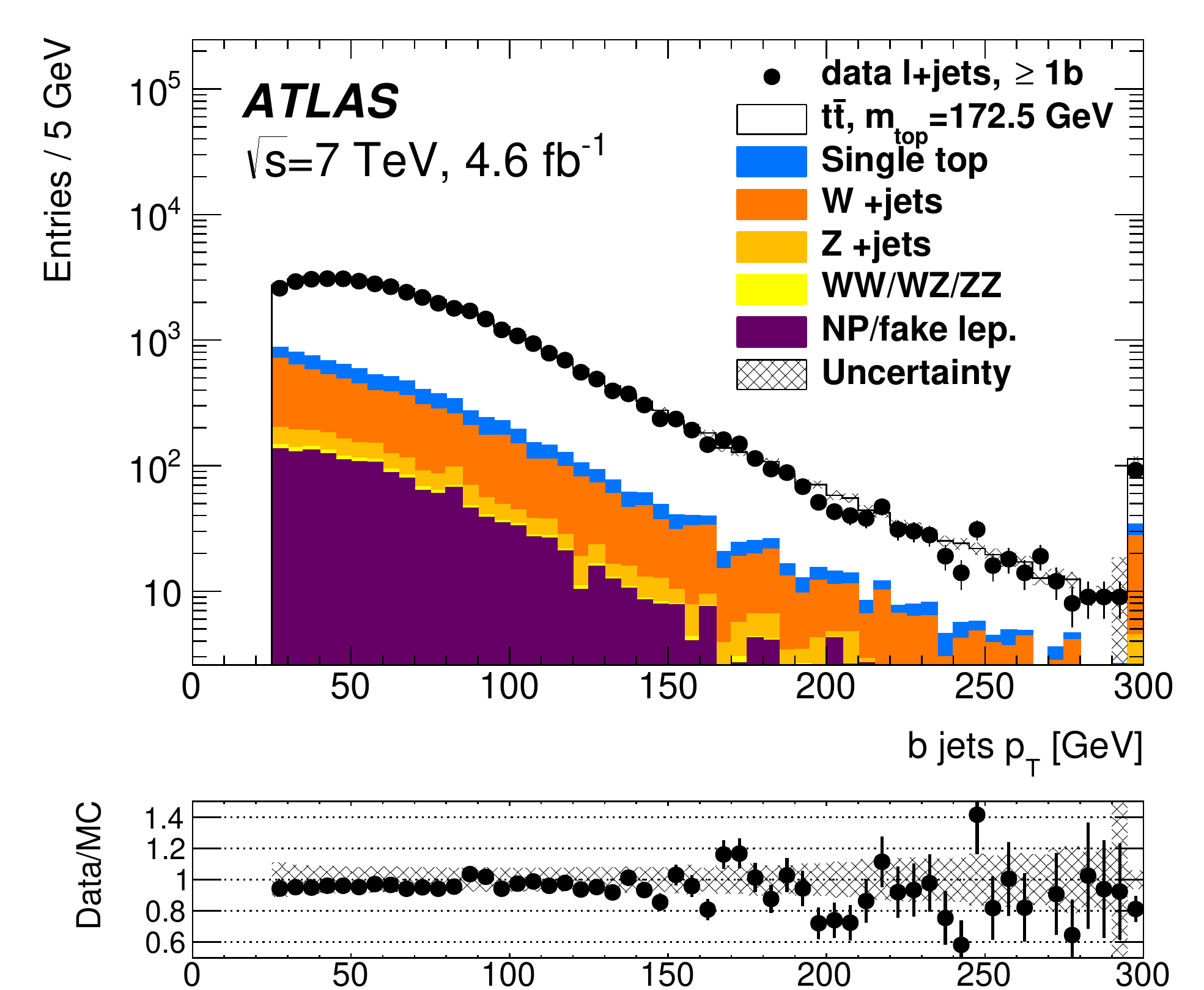}}
\subfloat[\ttbarll: \btagged\ jets]{
  \includegraphics[width=0.49\textwidth]{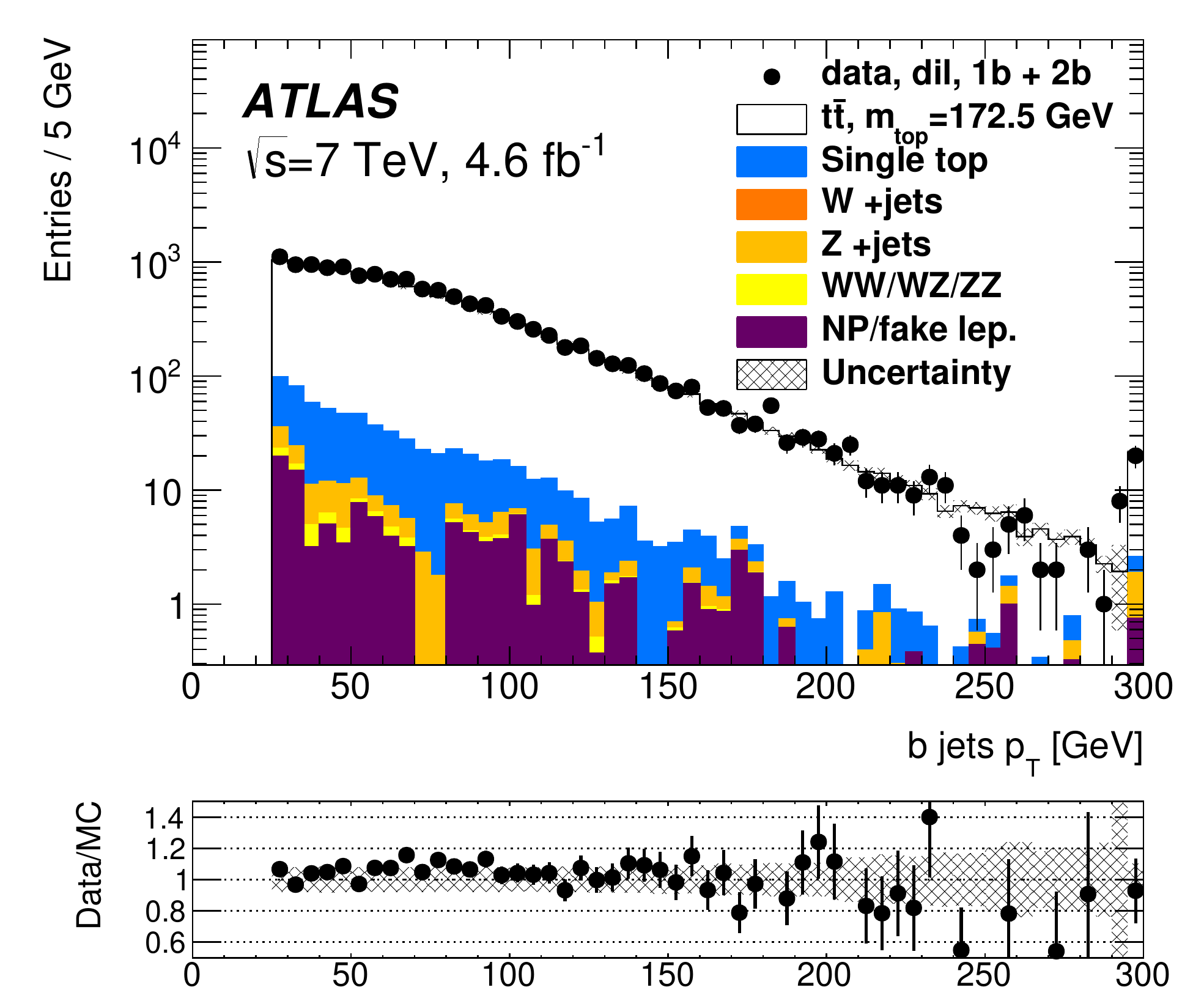}} \hfill
\caption{Distributions of the transverse momentum of the untagged and \btagged jets in the
  \ttbarlj\ analysis~(a, c) and of the charged lepton and \btagged\ jets \pt\ in
  the \ttbarll\ analysis~(b, d). The data are shown by the points and the
  signal-plus-background prediction by the solid histogram. The hatched area is
  the combined uncertainty on the prediction described in
  Sect.~\ref{sec:evtyields}, and the rightmost bin contains the overflow if
  present. For each figure, the ratio of the data to the MC prediction is also
  presented.
  \label{fig:rec}}
\end{figure*}

\section{Analysis method}
\label{sec:templates}
The observables exploited in the \mt\ analyses are: \mtr, \mWr, \rlbr\ in
the \ttbarlj\ channel and \mlbr\ in the \ttbarll\ channel.

In the \ttbarlj\ channel, templates of \mtr\ are constructed as a function of
the top quark mass used in the MC generation in the range 167.5--177.5~\GeV, in
steps of 2.5~\GeV. In addition, for the central mass point, templates of \mtr\
are constructed for an input value of the light-jet energy scale factor (\JSF)
in the range 0.95--1.05 in steps of 2.5\% and for an input value for the
relative $b$-to-light-jet energy scale factor (\bJSF) in the same range.
Independent MC samples are used for the different \mt\ mass points, and from
those samples templates with different values of \JSF\ and \bJSF\ are extracted
by appropriately scaling the four-momentum of the jets in each sample. The input
value for the \JSF\ is applied to all jets, whilst the input value for
the \bJSF\ is applied to all \bjets\ according to the information about the
generated quark flavour.
This scaling is performed after the various correction steps of the jet
calibration and before any event selection. This results in different events
entering the final selection from one energy scale variation to another.
Similarly, templates of \mWr\ are constructed as a function of an input \JSF\
combining the samples from all \mt\ mass points. Finally, templates of \rlbr\
are constructed as a function of \mt, and as a function of an input \bJSF\ at
the central mass point.

In the \ttbarll\ channel, signal templates for \mlbr\ are constructed as a
function of the top quark mass used in the MC generation in the range
167.5--177.5~\GeV, using separate samples for each of the five mass points.

The dependencies of the \mtr\ and \mlbr\ distributions on the underlying \mt\
used in the MC simulation are shown Fig.~\ref{fig:templtop}(a, b), for events
with at least (exactly) two \btagged\ jets, for the \ttbarlj\ (\ttbarll)
channel.
The \mtr\ and \mlbr\ distributions shown in Fig.~\ref{fig:templtop}(c--f),
exhibit sizeable sensitivity to global shifts of the \JSF\ and the \bJSF. These
effects introduce large systematic uncertainties on \mt\ originating from the
uncertainties on the JES and bJES, unless additional information is exploited.
As shown for the \ttbarlj\ channel in Fig.~\ref{fig:templaddvar}(a, c, e),
the \mWr\ distribution is sensitive to changes of the \JSF, while preserving its
shape under variations of the input \mt\ and \bJSF. As originally proposed in
Ref.~\cite{CDFlj2006}, a simultaneous fit to \mtr\ and \mWr\ is used to mitigate
the JES uncertainty.
The \rlbr\ distributions show substantial sensitivity to the \bJSF, 
and some dependence on the
assumed \mt\ in the simulation, Fig.~\ref{fig:templaddvar}(b, d,
f). Complementing the information carried by the \mtr\ and \mWr\
observables, \rlbr\ is used in an unbinned likelihood fit to the data to
simultaneously determine \mt,
\JSF, and \bJSF. The per-event correlations of any pair of observables
(\mtr, \mWr, and \rlbr) are found to be smaller than 0.15 and are neglected in
this procedure.
%
\begin{figure*}[tbp!]
\centering
\subfloat[\mtr, at least two \btagged\ jets]{
  \includegraphics[width=0.48\textwidth]{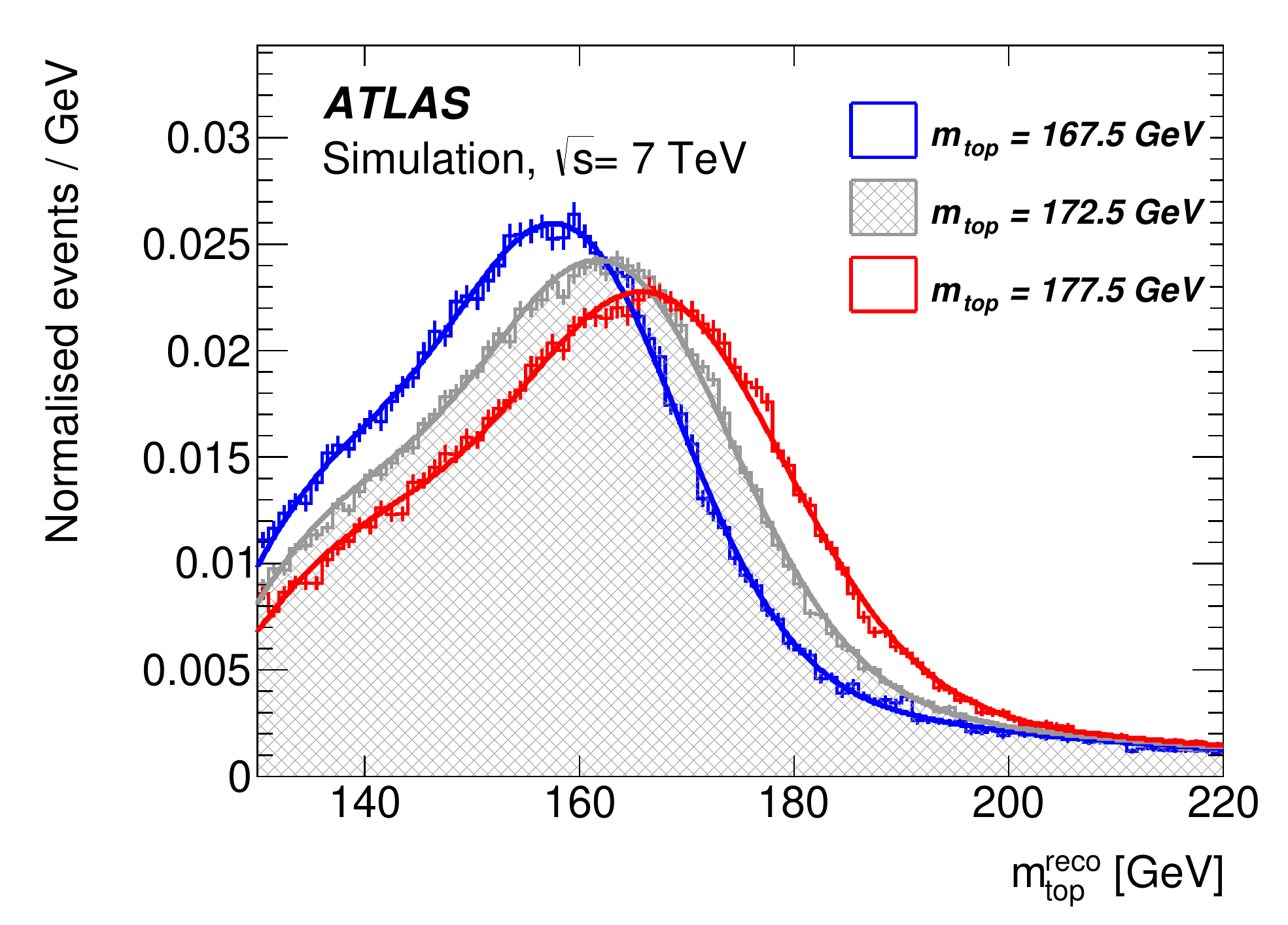}}
\subfloat[\mlbr, exactly two \btagged\ jets]{
  \includegraphics[width=0.48\textwidth]{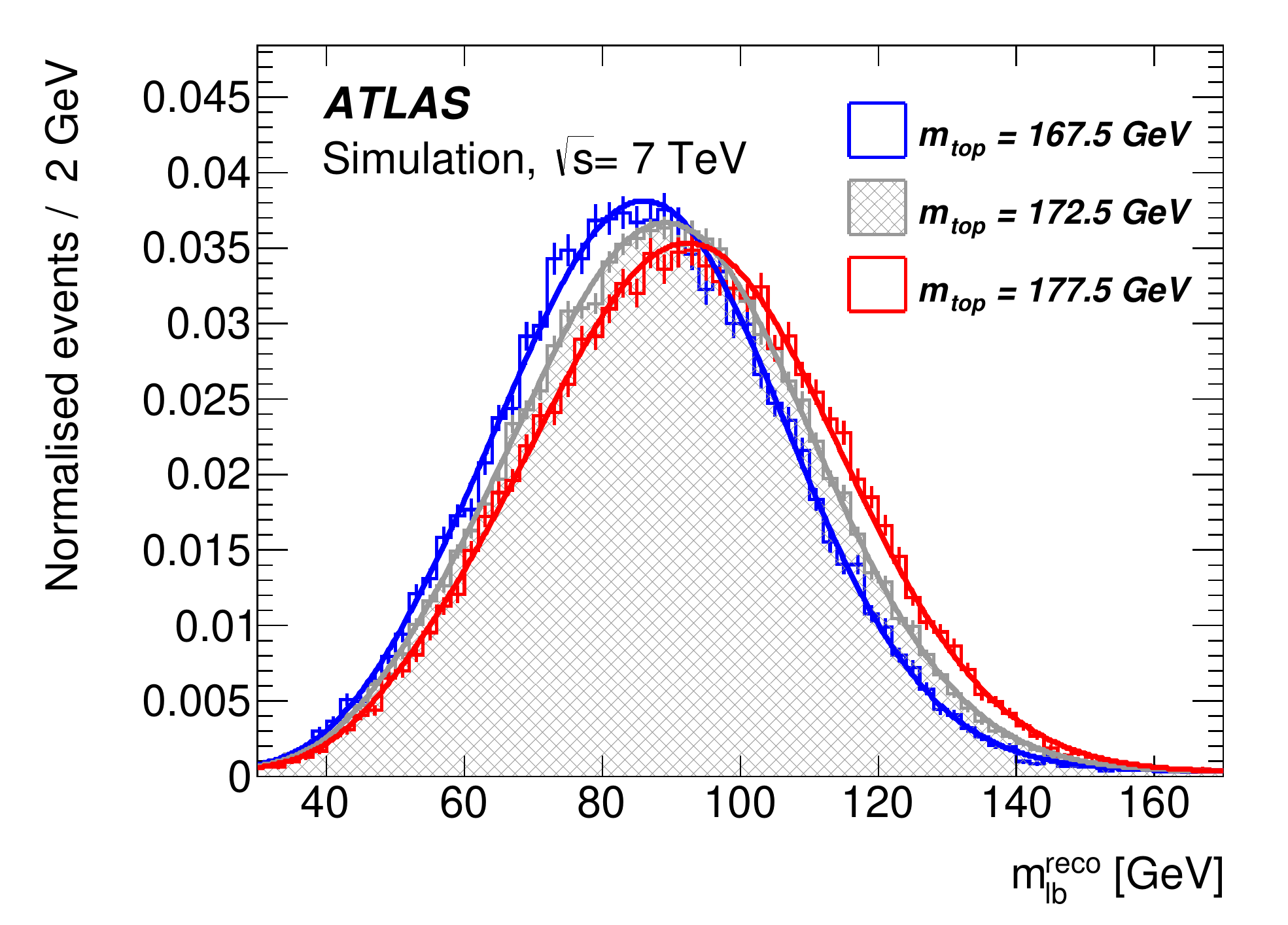}}
\hfill
\subfloat[\mtr, at least two \btagged\ jets]{
  \includegraphics[width=0.48\textwidth]{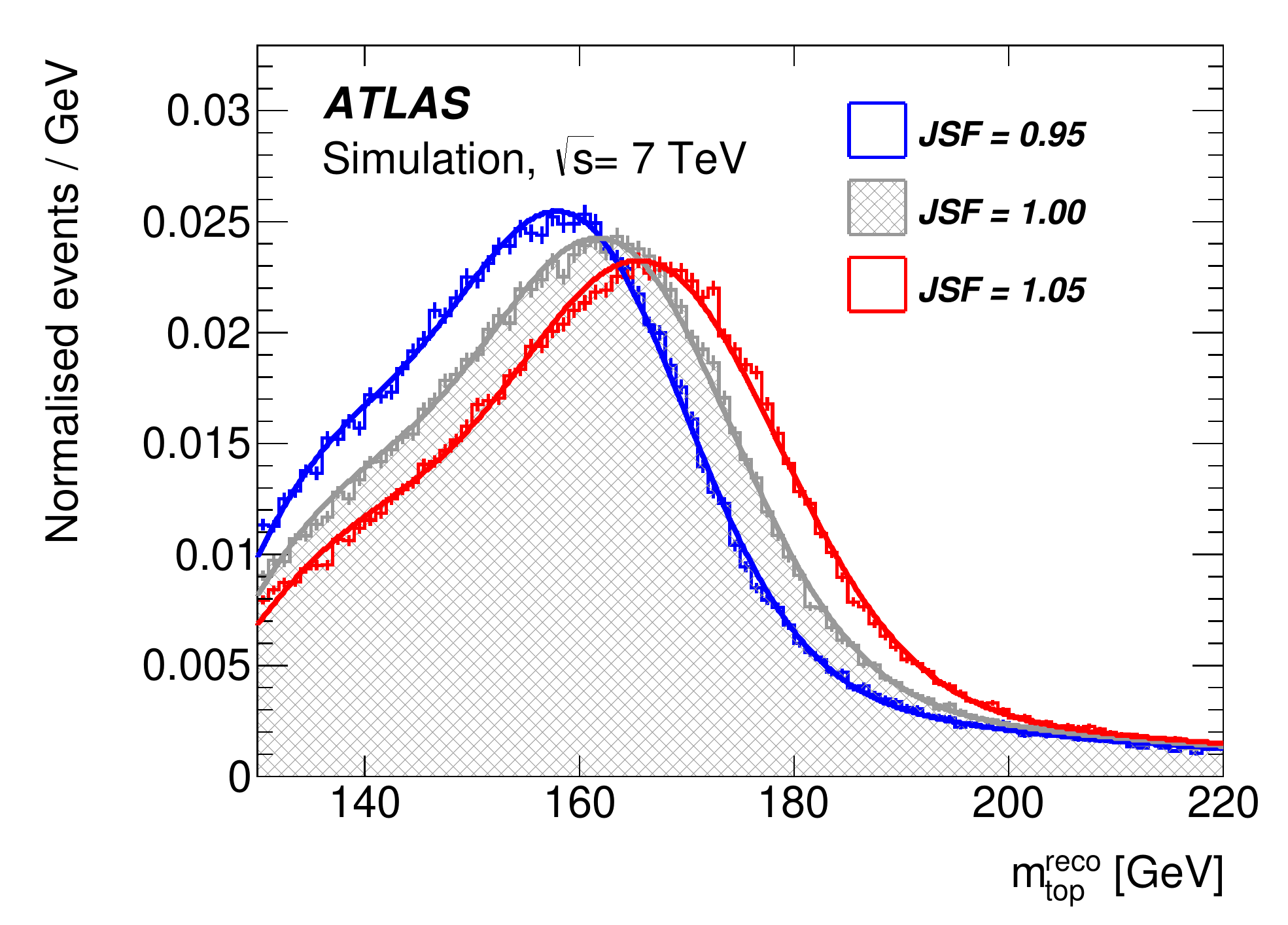}}
\subfloat[\mlbr, exactly two \btagged\ jets]{
  \includegraphics[width=0.48\textwidth]{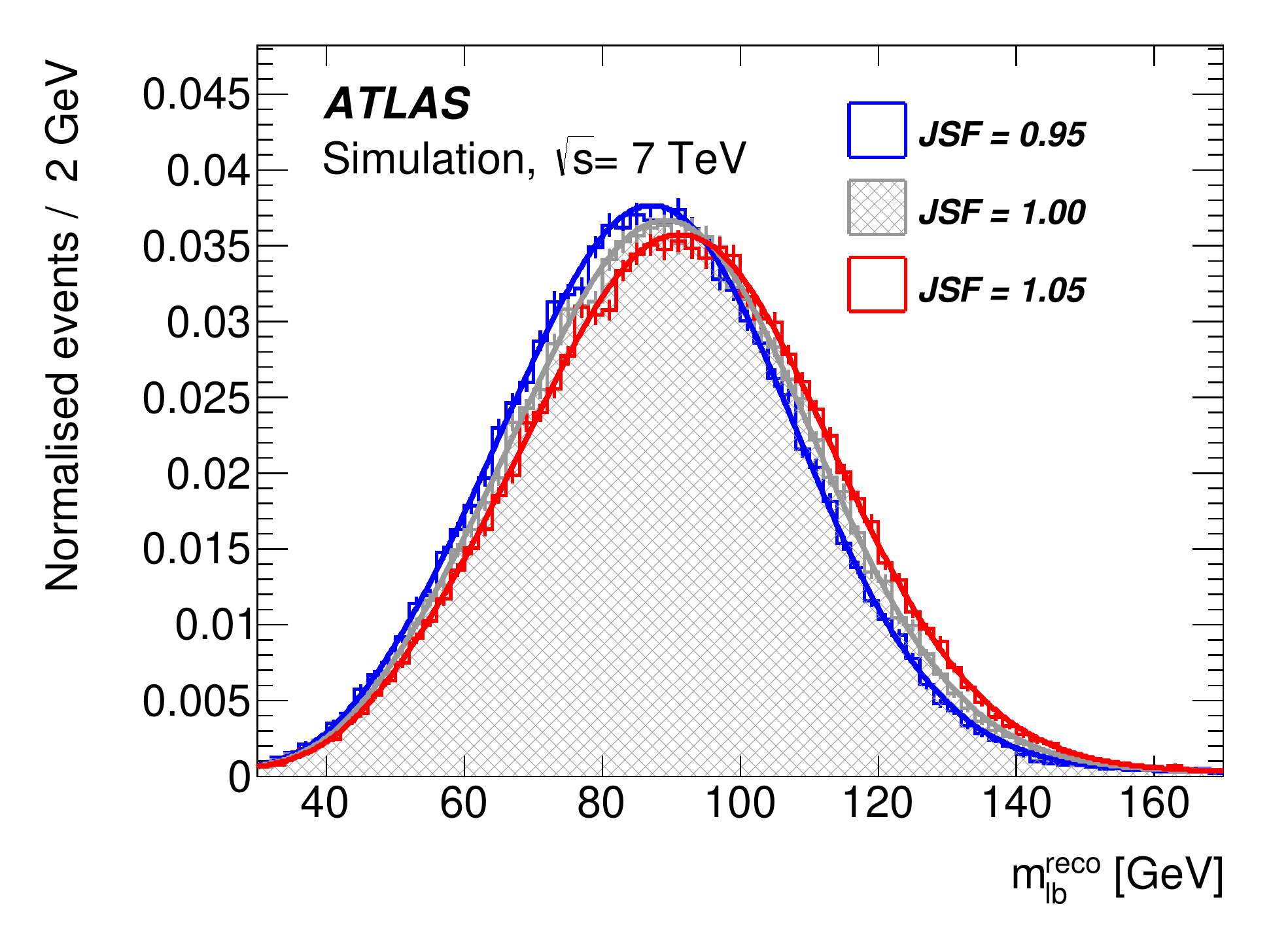}}
\hfill
\subfloat[\mtr, at least two \btagged\ jets]{
  \includegraphics[width=0.48\textwidth]{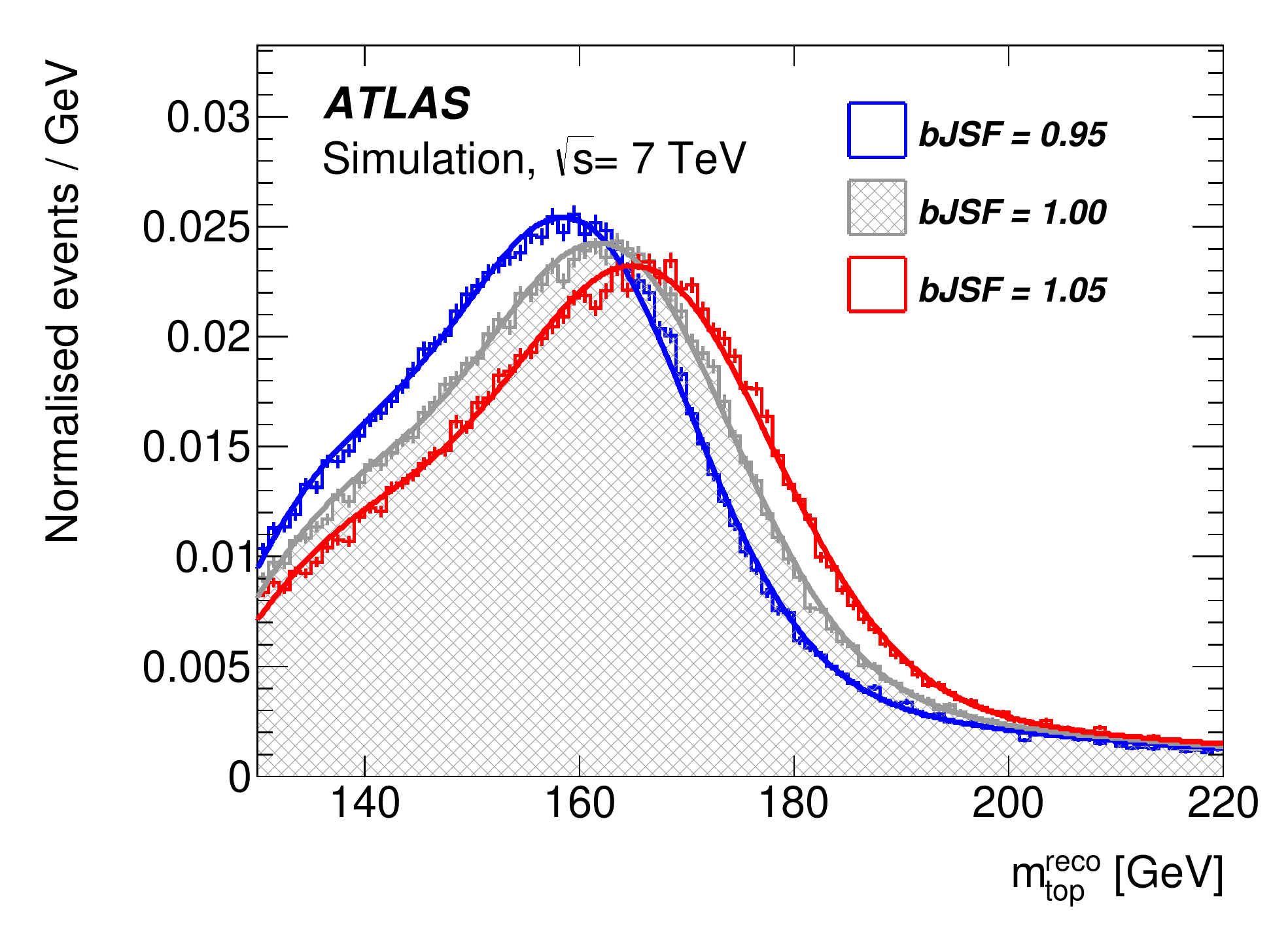}}
\subfloat[\mlbr, exactly two \btagged\ jets]{
  \includegraphics[width=0.48\textwidth]{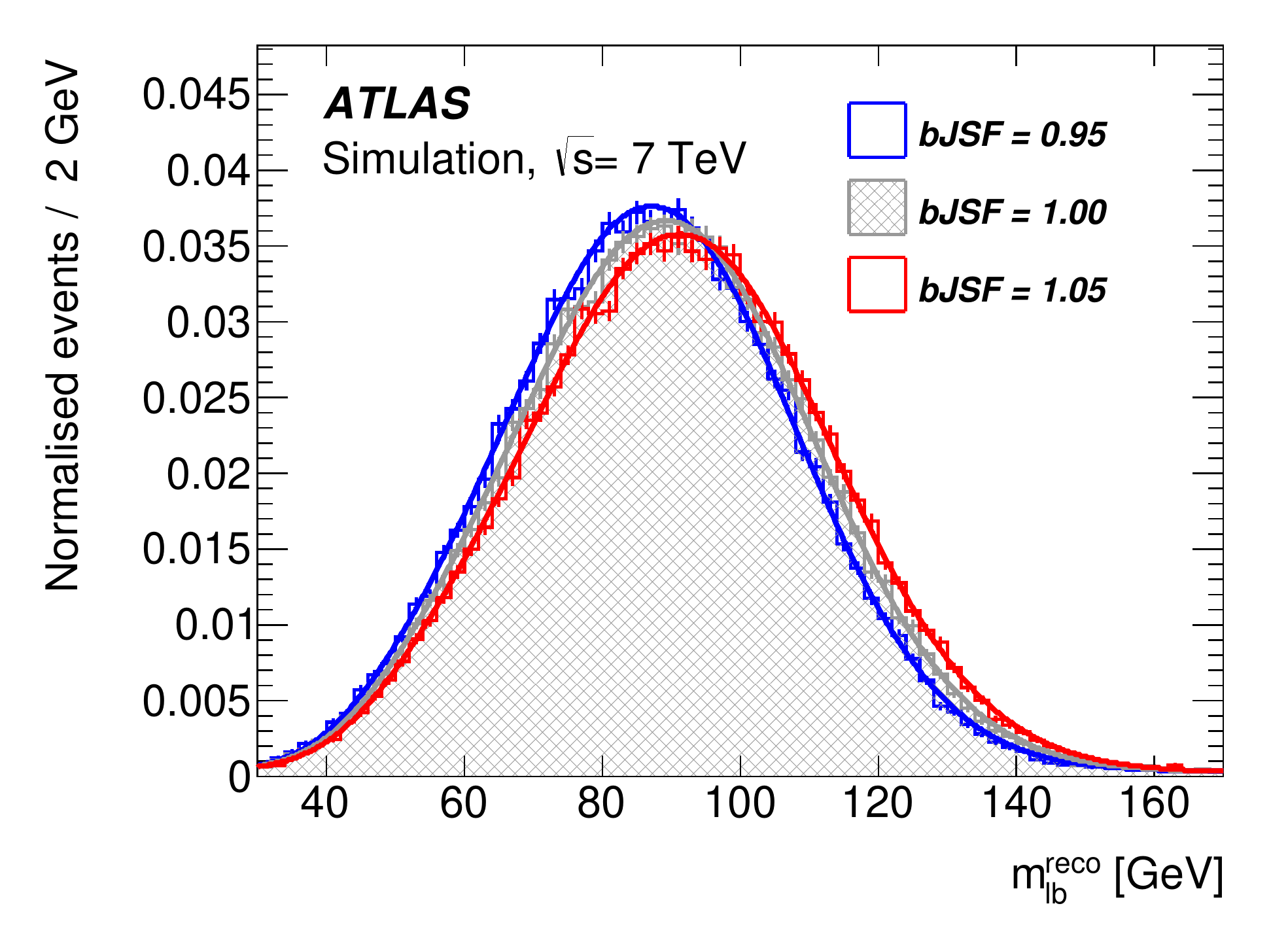}}
\caption{Distributions of \mtr\ in the \ttbarlj\ channel~(left)
  and \mlbr\ in the \ttbarll channel~(right) and their template
  parameterisations for the signal, composed of simulated \ttbar\ and single top
  quark production events. The expected sensitivities of \mtr\ and \mlbr\ are
  shown for events with at least two (or exactly two) \btagged\ jets. Figures~(a,
  b) report the distributions for different values of the input \mt\ (167.5,
  172.5 and 177.5 \GeV). Figures~(c, d) and (e, f) show the \mtr\ and \mlbr\
  distribution for \mt=172.5~\GeV, obtained with \JSF\ or \bJSF\ of 0.95, 1.00
  and 1.05, respectively. 
  Each distribution is overlaid with the corresponding probability density
  function that is obtained from the combined fit to all signal templates for
  all abservables.
  \label{fig:templtop}}
\end{figure*}
%
\begin{figure*}[tbp!]
\centering
\subfloat[\mWr, at least two \btagged\ jets]{
  \includegraphics[width=0.48\textwidth]{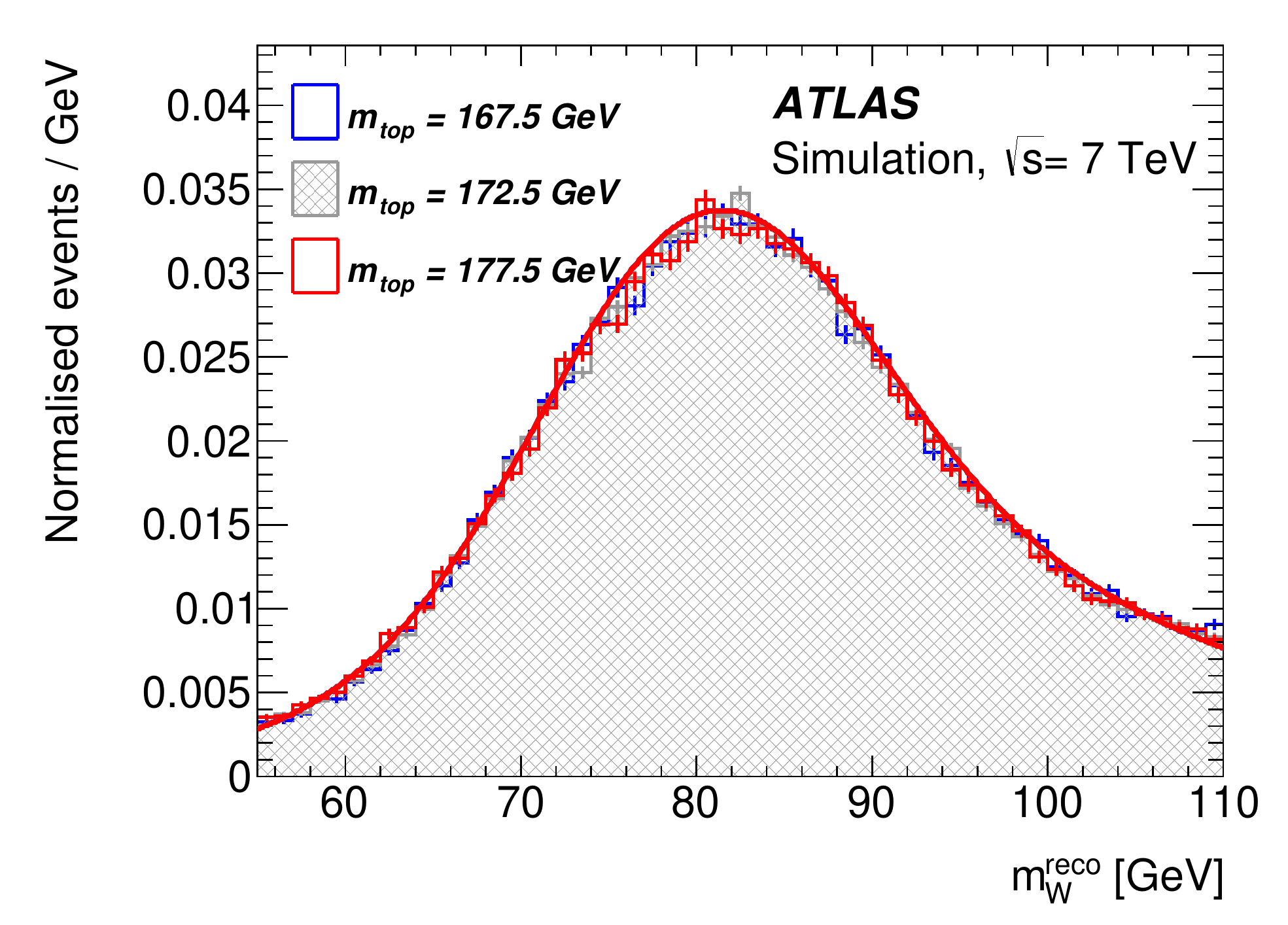}}
\subfloat[\rlbr, at least two \btagged\ jets]{
  \includegraphics[width=0.48\textwidth]{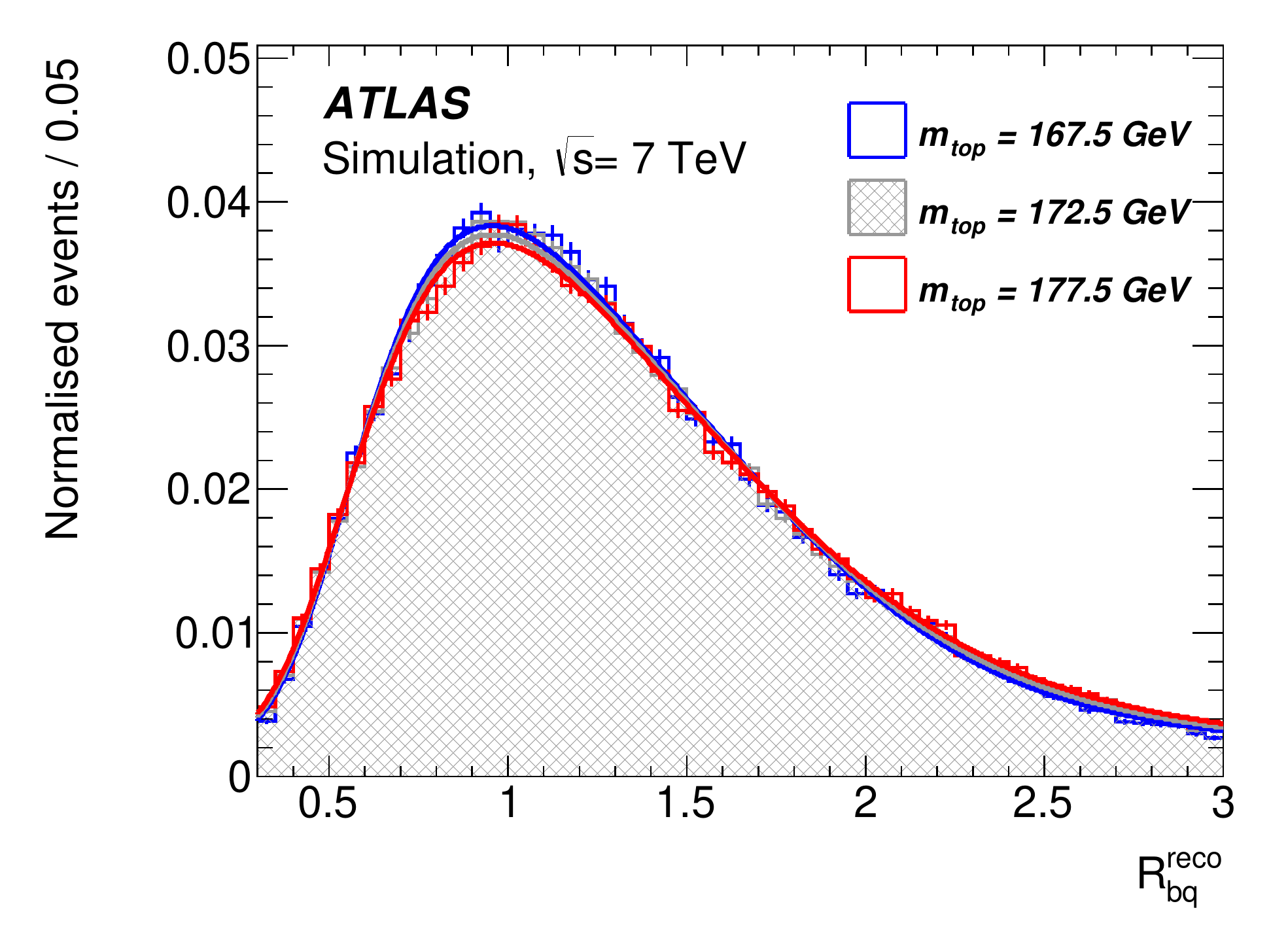}}
\hfill
\subfloat[\mWr, at least two \btagged\ jets]{
  \includegraphics[width=0.48\textwidth]{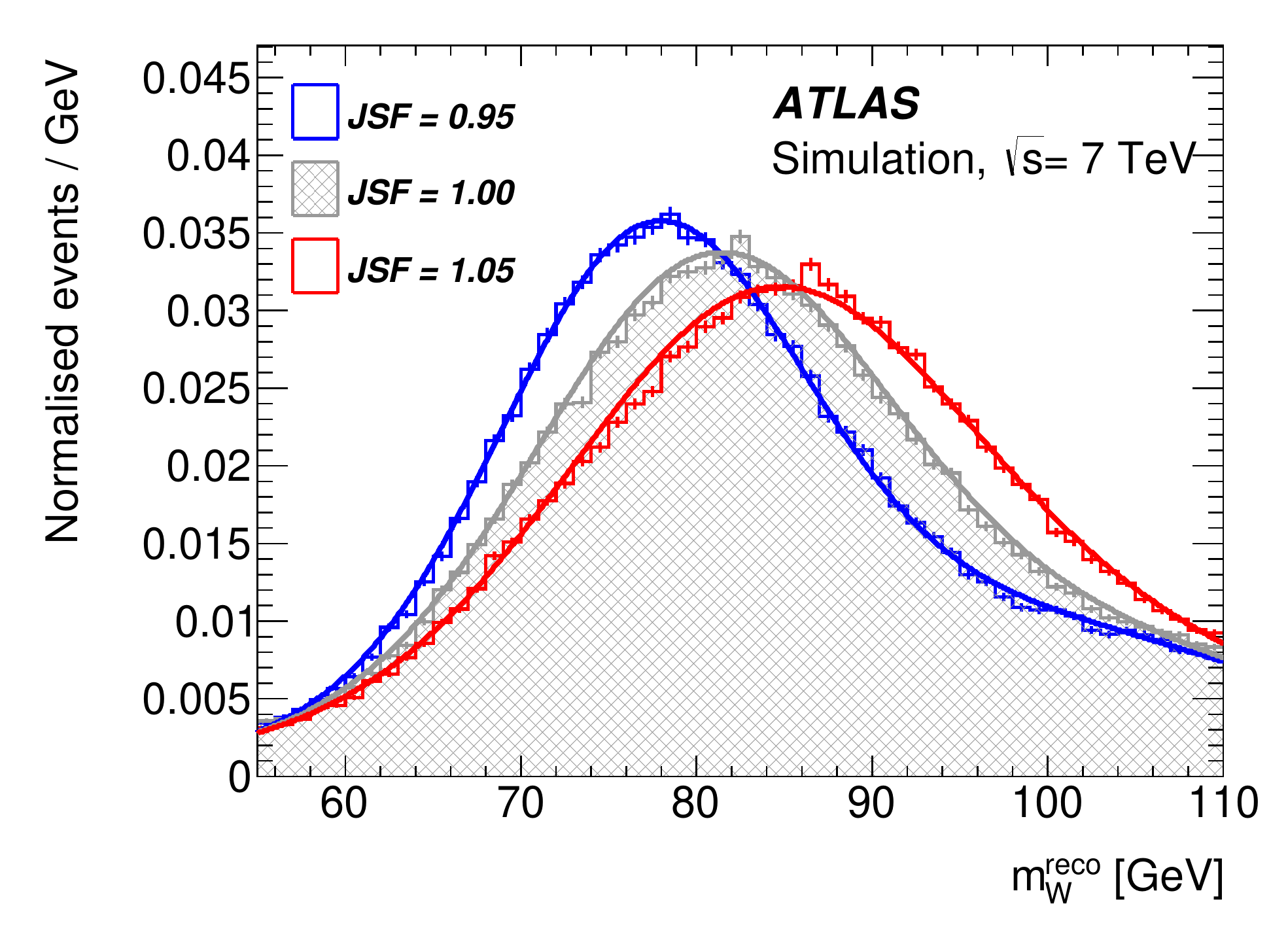}}
\subfloat[\rlbr, at least two \btagged\ jets]{
  \includegraphics[width=0.48\textwidth]{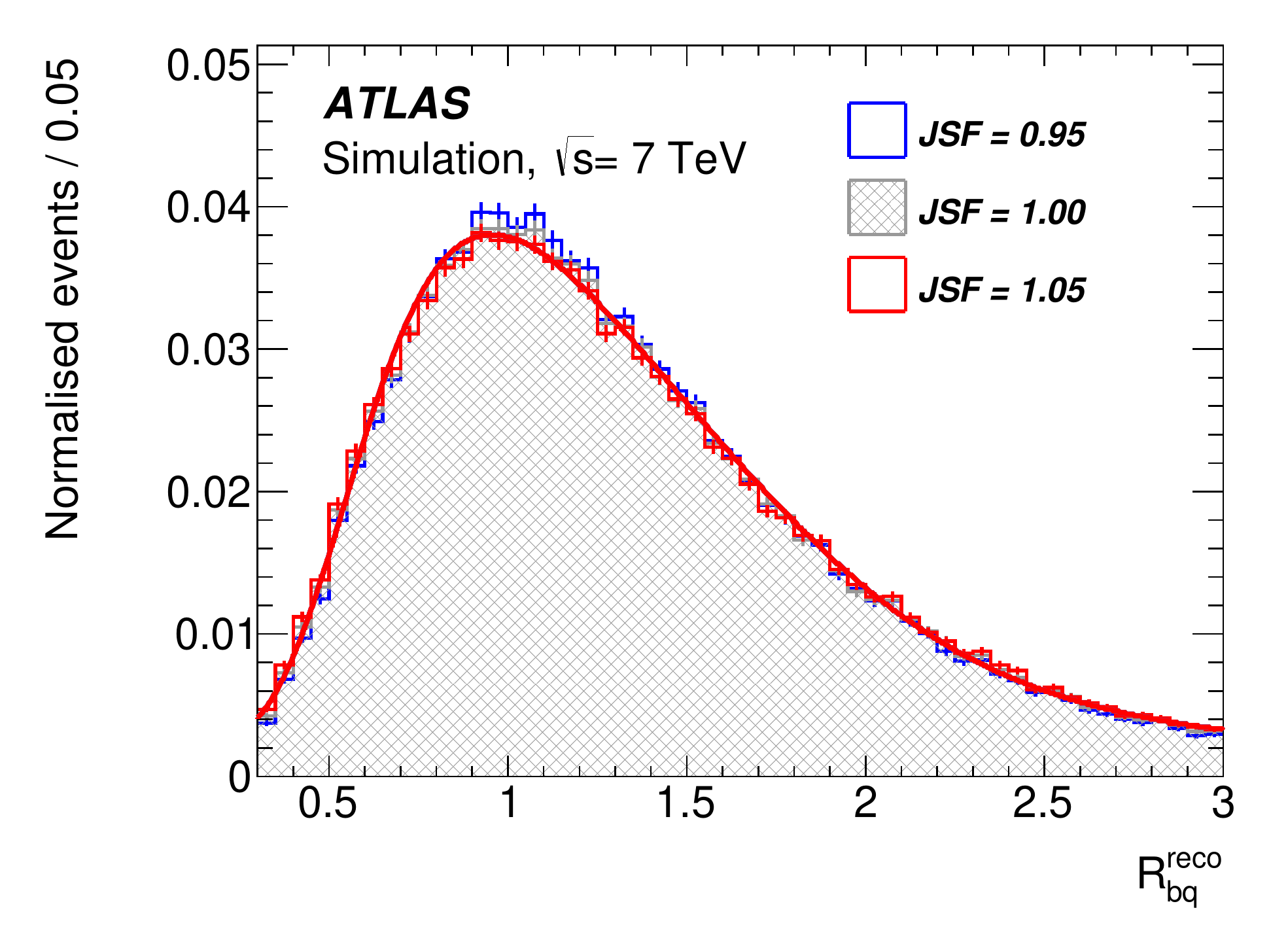}}
\hfill
\subfloat[\mWr, at least two \btagged\ jets]{
  \includegraphics[width=0.48\textwidth]{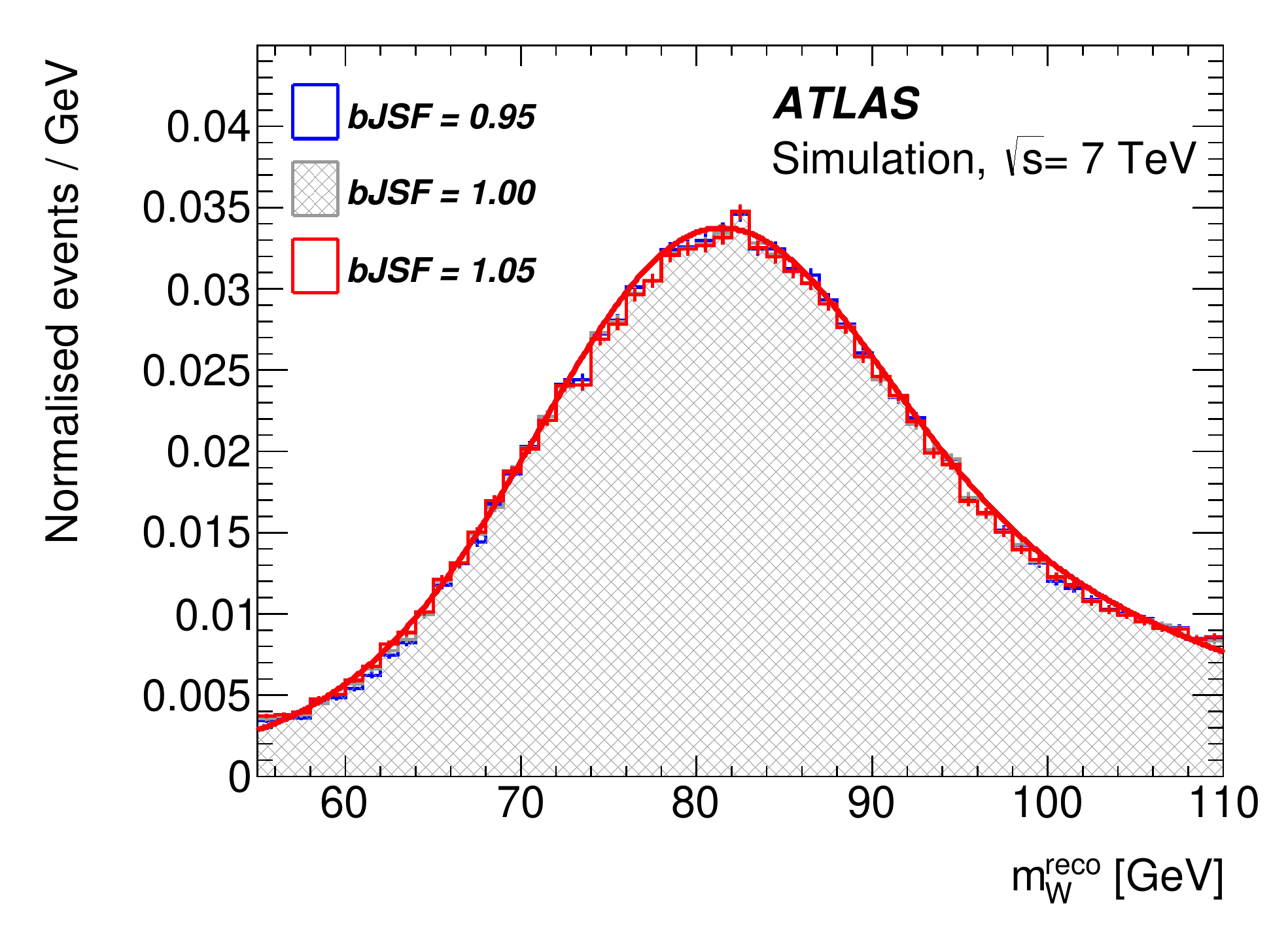}}
\subfloat[\rlbr, at least two \btagged\ jets]{
  \includegraphics[width=0.48\textwidth]{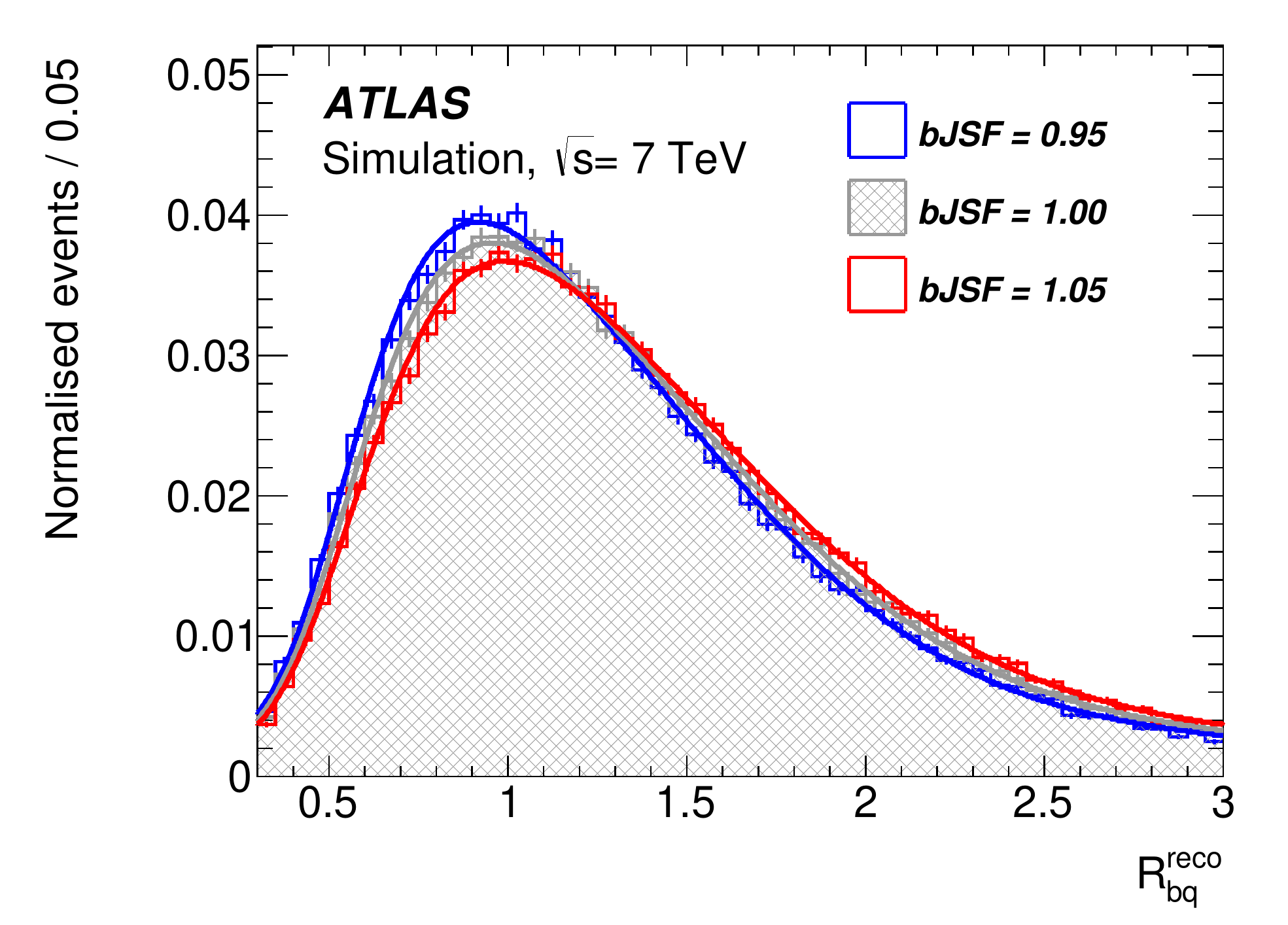}}
\caption{Distributions of \mWr\ (left) and \rlbr\ (right) in the \ttbarlj\ channel 
  and their template parameterisations for the signal, composed of
  simulated \ttbar\ and single top quark production events. The expected
  sensitivity of \mWr\ and \rlbr\ are shown for events with at least
  two \btagged\ jets. Figures~(a, b) report the distributions for different values
  of the input \mt\ (167.5, 172.5 and 177.5 \GeV). Figures~(c, d) and (e, f) show
  the \mWr\ and \rlbr\ distribution for \mt=172.5~\GeV, obtained with \JSF\
  or \bJSF\ of 0.95, 1.00 and 1.05, respectively.
  Each distribution is overlaid with the corresponding probability density
  function that is obtained from the combined fit to all signal templates for
  all abservables.
  \label{fig:templaddvar}}
\end{figure*}
%
\subsection{Templates and fits in the \ttbarlj\ channel}
Signal templates are derived for the three observables for all \mt-dependent
samples, consisting of the \ttbar\ signal events, together with single top quark
production events.
The signal templates for the \mtr, \mWr\ and \rlbr\ distributions are fitted to
the sum of a Gaussian function and a Landau function for \mtr\ and
\rlbr, and to a sum of two Gaussian functions for \mWr\
(Figs.~\ref{fig:templtop} and~\ref{fig:templaddvar}).
For the background, the \mtr\ distribution is fitted to a Landau function, while
both the \mWr\ and the \rlbr\ distributions are fitted to the sum of two
Gaussian functions.
To exploit the different sensitivities to the underlying \mt, \JSF\ and \bJSF,
all template fits are performed separately for events with one
\btagged\ jet, and for events with at least two \btagged\ jets.

From individual fits to all signal templates listed above, it was verified that
the parameters of the fitting functions depend linearly on the respective
parameter \mt, \JSF\ or \bJSF. Consequently, this linearity is imposed when
parametrising the fitting functions for the combined fit to all signal templates
for the three observables. 
For the signal, the parameters of the fitting functions for \mtr\ depend
linearly on \mt, \JSF\ and \bJSF. The parameters of the fitting functions
of \mWr\ depend linearly on the \JSF. Finally, the parameters of the fitting
functions of \rlbr\ depend linearly on the \bJSF and on
\mt. As shown in Fig.~\ref{fig:templaddvar}, the dependencies of
\mWr\ on \mt\ and \bJSF, and of \rlbr\ on \JSF\ are negligible.
For the background, the parameter dependencies of the fitting functions are the
same except that, by construction, they do not depend on \mt.

Signal and background probability density functions \Psig\ and
\Pbkg\ for the \mtr, \mWr\ and \rlbr\ distributions are used in an
unbinned likelihood fit to the data for all events, $i=1,\dots N$.
The likelihood function maximised is:
%
\begin{eqnarray}
 \Likeljets(\mt, \JSF, \bJSF, \fbkg)  &=& 
 \prod_{i=1}^{N} \Ptop(\mtri\,\vert\,\mt, \JSF, \bJSF, \fbkg) \nonumber \\ 
 &&\quad\quad\times\,\PW(\mWri\,\vert\,\JSF, \fbkg) \nonumber \\ 
 &&\quad\quad\times\,\Prlb(\rlbri\,\vert\,\mt,\bJSF, \fbkg), 
\label{eq:LikeLJ} 
\end{eqnarray}
\noindent
with:
\begin{eqnarray*}
  \Ptop(\mtri\,\vert\,\mt, \JSF, \bJSF, \fbkg) &=& 
 (1-\fbkg)\cdot\Ptopsig(\mtri\,\vert\,\mt, \JSF, \bJSF) +
 \nonumber \\&&\quad\quad\,
 \fbkg\cdot\Ptopbkg(\mtri\,\vert\,\JSF, \bJSF)\,\,,
 \nonumber \\
   \PW(\mWri\,\vert\,\JSF, \fbkg) &=& (1-\fbkg)\cdot\PWsig(\mWri\,\vert\,\JSF) +
 \nonumber \\
       && \quad\quad\, \fbkg\cdot\PWbkg(\mWri\,\vert\,\JSF)\,\,,
 \nonumber \\
   \Prlb(\rlbri\,\vert\,\mt,\bJSF, \fbkg)
  &=& (1-\fbkg)\cdot\Prlbsig(\rlbri\,\vert\,\mt,\bJSF) +
 \nonumber \\
   && \quad\quad\, \fbkg\cdot\Prlbbkg(\rlbri\,\vert\,\bJSF)\,\,
 \nonumber
\end{eqnarray*}
%
where the fraction of background events is denoted by \fbkg. 
The parameters to be determined by the fit are \mt, \JSF, \bJSF\ and
\fbkg, where \fbkg\ is determined separately for the \ttbarlj\ data sets
with exactly one or at least two \btagged\ jets.

Pseudo-experiments are used to verify the internal consistency of the fitting
procedure and to obtain the expected statistical uncertainty corresponding to a
data sample of \atlumo~\ifb. For each choice of the input parameters,
500~pseudo-experiments are generated.
To retain the correlation of the analysis observables, individual MC events
drawn from the full simulated event samples are used, rather than sampling from
the separate \mtr, \mWr, and \rlbr\ distributions.
For all five parameters, good linearity is found between the input parameters
used to perform the pseudo-experiments, and the results of the fits.
Within their statistical uncertainties, the mean values and widths of the pull
distributions are consistent with the expectations of zero and one,
respectively. This means the method is unbiased with appropriate statistical
uncertainties.
The expected statistical uncertainties on \mt\ including the statistical
contributions from the simultaneous fit of the \JSF and \bJSF\ obtained from
pseudo-experiments at an input top quark mass of $\mt=172.5$~\GeV, and for a
luminosity of \atlumo~\ifb, are
\mtstaobps~\GeV\ and \mtstatbps~\GeV\ for the case of one \btagged\
jet\ and for the case of at least two \btagged\ jets,
respectively. The results correspond to the mean value and the
standard deviation of the distribution of the statistical uncertainties
of the fitted masses from the pseudo-experiments.
The different expected statistical uncertainties on \mt\ for the samples with
different numbers of \btagged\ jets, which are obtained from samples containing
similar numbers of events (see Table~\ref{tab:LJDLcutflow}), are mainly a
consequence of the different resolution on \mt.
%
\subsection{Templates and fits in the \ttbarll\ channel}
The signal \mlbr\ templates comprise both the \ttbar\ and the single top quark
production processes, and are fitted to the sum of a Gaussian function and a
Landau function, while the background distribution is fitted to a Landau
function.
Similarly to the \ttbarlj\ channel, all template fits are performed separately
for events with one \btagged\ jet, and for events with exactly two \btagged\
jets.
In Fig.~\ref{fig:templtop}(b) the sensitivity of the \mlbr\ observable to the
input value of the top quark mass is shown for the events with exactly
two \btagged\ jets, by the superposition of the signal templates and their fits
for three input \mt\ values.
For the signal templates, the parameters of the fitting functions of
\mlbr\ depend linearly on \mt.

Signal and background probability density functions for the \mlbr\ estimator are
built, and used in an unbinned likelihood fit to the data for all events,
$i=1,\dots N$.  The likelihood function maximised is:
%
\begin{equation}
\Likedil(\mt, \fbkg) = 
\prod_{i=1}^{N} 
\left[ (1-\fbkg)\cdot \Ptopsig(\mlbi \,\vert\, \mt) + \fbkg \cdot \Ptopbkg(\mlbi) \right],
\label{eq:LikeDL}
\end{equation}
%
where, as for the \ttbarlj\ case, \Ptopsig\ and \Ptopbkg\ are the signal and
background probability density functions and \fbkg\ is the fraction of
background events in the selected data set.

Using pseudo-experiments, also for this decay channel good linearity is found
between the input top quark mass used to perform the pseudo-experiments, and the
results of the fits.
Within their statistical uncertainties, the mean values and widths of the pull
distributions are consistent with the expectations of zero and one,
respectively.
The expected statistical uncertainties on \mt\ obtained from pseudo-experiments
for an input top quark mass of $\mt=172.5$~\GeV, and for a luminosity
of \atlumo~\ifb, are \mtstaobdlps~\GeV\ and
\mtstatbdlps~\GeV\ for events with exactly one or two \btagged\ jets,
respectively.
As for the \ljets\ channel, the different expected statistical uncertainties on
\mt\ for the samples with different numbers of \btagged\ jets, which are obtained
from samples containing similar numbers of events (see
Table~\ref{tab:LJDLcutflow}), are mainly a consequence of the different
resolution on \mt.
%
\subsection{Combined likelihood fit to the event samples} 
The final results for both the \ljets\ and \dil\ final states are obtained
combining at the likelihood level the events with one or more
\btagged\ jets.
The measured \mt\ is assumed to be the same in these two sub-samples per decay
channel. Similarly, the \JSF\ and the \bJSF\ are taken to be the same for the
samples of the \ttbarlj\ analysis with different \btagged\ jet multiplicities.
On the contrary, the background fractions for the two decay channels, and for
the samples with different numbers of \btagged\ jets, are kept independent,
corresponding to four individual parameters ($f_{\rm bkg}^{\ell+{\rm jets}, 1b}$, $f_{\rm
bkg}^{\ell+{\rm jets}, 2b}$, $f_{\rm bkg}^{{\rm dil}, 1b}$, $f_{\rm bkg}^{{\rm dil}, 2b}$).

The combined likelihood fit allows the statistical uncertainties on the fitted
parameters to be reduced, while mitigating some systematic effects.
The expected statistical precision on \mt, for an input top quark mass of
$\mt=172.5$~\GeV, a luminosity of \atlumo~\ifb, and in the combined one or
more \btagged\ jets event sample, is $0.76 \pm 0.01 \GeV$ and $0.54 \pm
0.01 \GeV$ for the \ttbarlj\ and \ttbarll\ analyses, respectively.
%
\section{Top quark mass measurements}
\label{sec:result}
The results of the fits for the \ttbarlj\ and \ttbarll\ analyses are:
%
\begin{eqnarray*}
\label{Eq:resultstat}
 \mtlj & = & \mtval \pm \mtstaScales~\mathrm{(stat + \JSF + \bJSF)}~\GeV,\\
 \JSF  & = & \XZst{\jsfval}{\jsfsta},\\
 \bJSF & = & \XZst{\bjsfval}{\bjsfsta}, \\
 \mtdl & = & \XZst{\mtdlval}{\mtdlsta}~\GeV.
\end{eqnarray*}
%
For the \ttbarlj\ channel, the fitted background fractions amount to
$18.4\% \pm 2.2\%$ and $2.4\%\pm 1.5\%$ for one \btagged\ jet and the at least
two \btagged\ jets samples respectively.
The corresponding values for the \ttbarll\ analysis are $3.5\% \pm 3.7\%$ and
$1.4\%\pm 2.2\%$ for one \btagged\ jet and the two \btagged\ jets samples
respectively. All quoted uncertainties are statistical only.
These fractions are consistent with the expectations given in
Table~\ref{tab:LJDLcutflow}.
The correlation matrices for the fitted parameters in the \ttbarlj\
and \ttbarll\ analyses are reported in Table~\ref{tab:paramcorr}.
%
\begin{table}[!tp]
\parbox{.45\linewidth}{
\begin{center}
\footnotesize 
\begin{tabular}{|r||rrrrr|}
\hline
& \mtlj & \JSF & \bJSF & $f_{\rm bkg}^{\ell+{\rm jets}, 1b}$ & $f_{\rm bkg}^{\ell+{\rm jets}, 2b}$ \\ \hline
\mtlj                  &   1.00   &        &        &       & \\           
\JSF                   & --0.36   &   1.00 &        &       & \\           
\bJSF                  & --0.89   &   0.03 &  1.00  &       & \\           
$f_{\rm bkg}^{\ell+{\rm jets}, 1b}$  & --0.03   & --0.01 &  0.06  &  1.00 & \\  
$f_{\rm bkg}^{\ell+{\rm jets}, 2b}$  & --0.06   & --0.09 &  0.09  &  0.01 &  1.00    \\  \hline
\end{tabular}
\end{center}
}\hfill
\parbox{.45\linewidth}{
\begin{center}
\footnotesize 
\begin{tabular}{|r||rrr|}
\hline
& \mtdl & $f_{\rm bkg}^{{\rm dil}, 1b}$ & $f_{\rm bkg}^{{\rm dil}, 2b}$ \\ \hline
\mtdl                     &  1.00 &       &         \\  
 $f_{\rm bkg}^{{\rm dil}, 1b}$  &  0.07 &  1.00 &         \\
 $f_{\rm bkg}^{{\rm dil}, 2b}$  & --0.14 & --0.01&  1.00    \\  \hline
\end{tabular}
\end{center}
}
\caption{The correlations of the fitted parameters used in the likelihood 
maximisation of the \ttbarlj\ analysis (left) and the \ttbarll\ analysis
(right).}
\label{tab:paramcorr}
\end{table}

Figure~\ref{Fig:mjjbdata} shows the \mtr, \mWr, \rlbr\ and \mlbr\ distributions
in the data together with the corresponding fitted probability density functions
for the background alone and for the sum of signal and background.
The uncertainty bands are obtained by varying the three fitted parameters \mt,
\JSF, and \bJSF\ within $\pm 1\sigma$~of their full uncertainties
taking into account their correlation, while keeping the background fractions
fixed.
The individual systematic uncertainties and the correlations are discussed in
Sect.~\ref{sec:syst} and Sect.~\ref{sec:resultcomb}, respectively.
The band shown is the envelope of all probability density functions obtained
from 500 pseudo-experiments varying the parameters.
Within this band, the data are well described by the fitted probability density
function.
%
\begin{figure*}[bp!]
\centering
\subfloat[\mtr, \ljets\ ]{
  \includegraphics[width=0.49\textwidth]{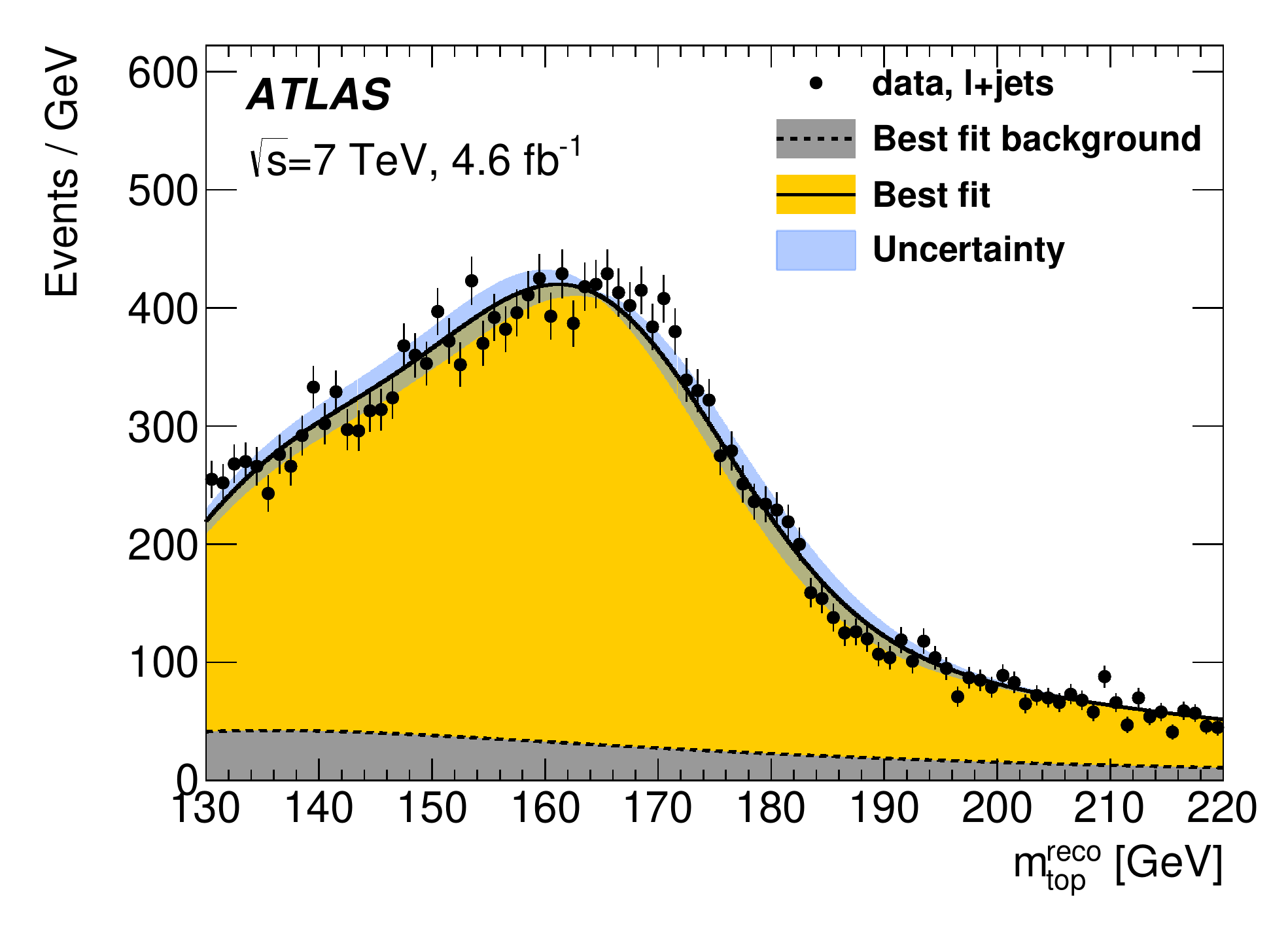}}
\subfloat[\mWr, \ljets\ ]{
  \includegraphics[width=0.49\textwidth]{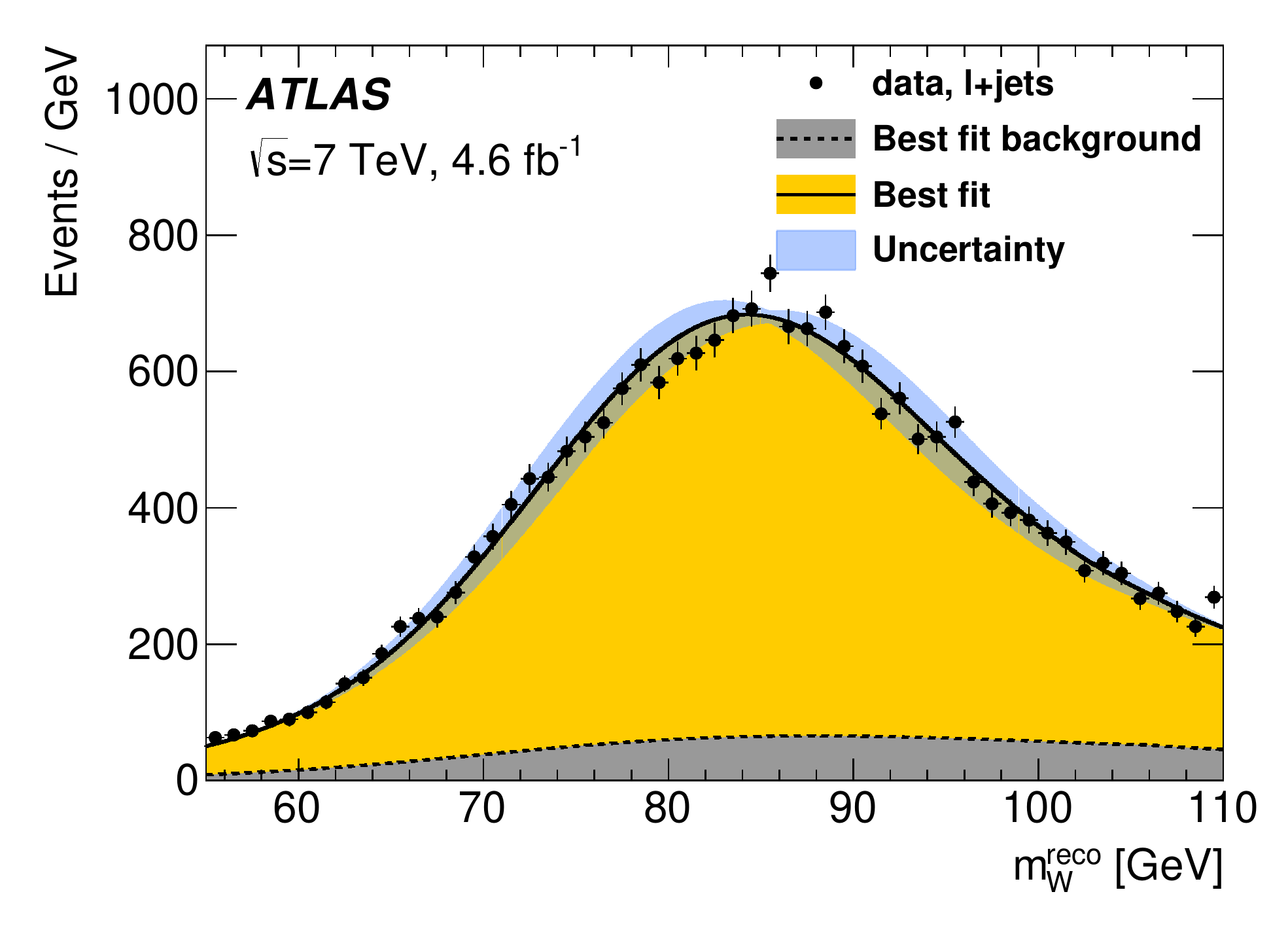}}
\hfill
\subfloat[\rlbr, \ljets\ ]{
  \includegraphics[width=0.49\textwidth]{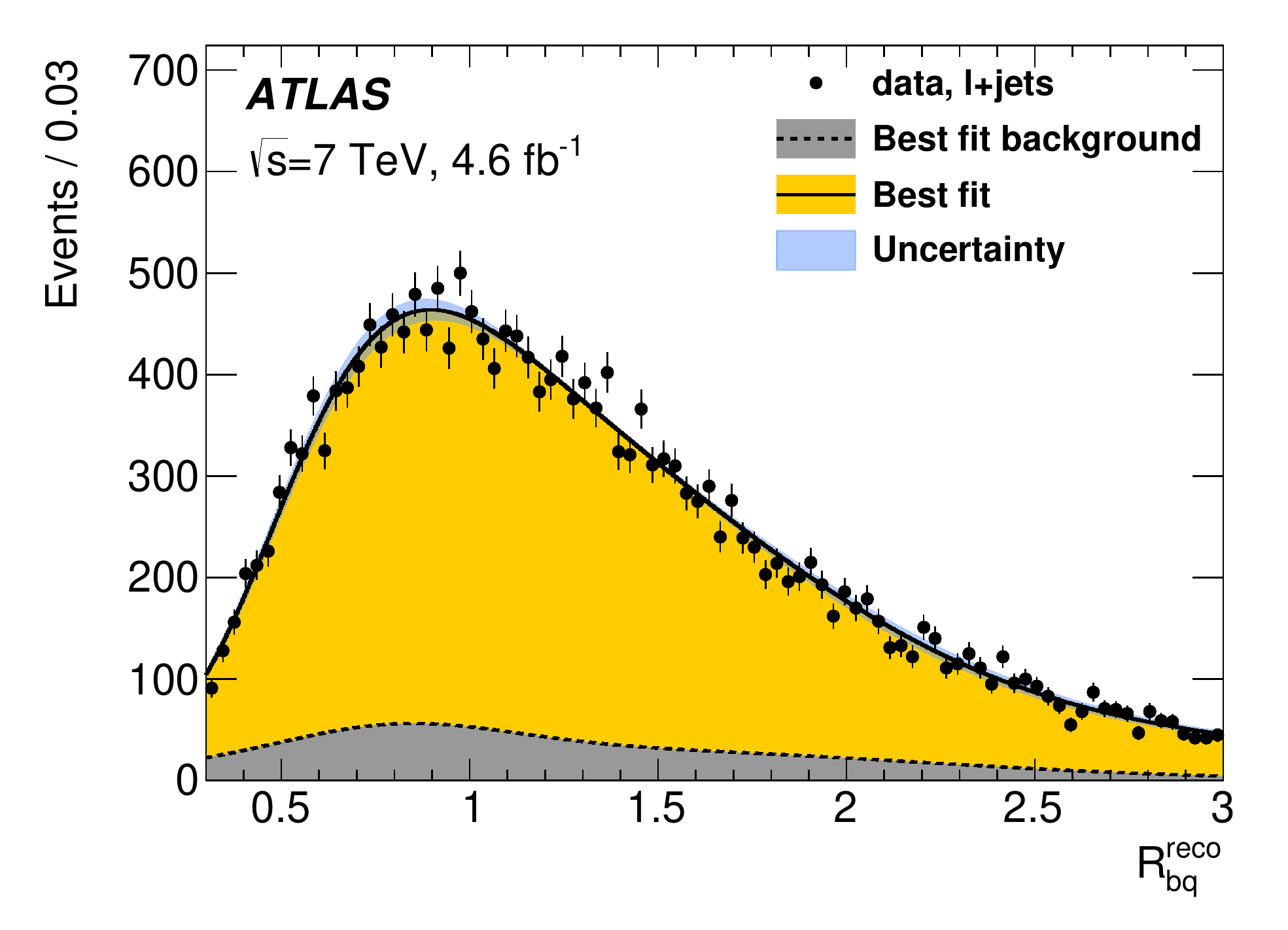}}
\subfloat[\mlbr, \dil\ ]{
  \includegraphics[width=0.49\textwidth]{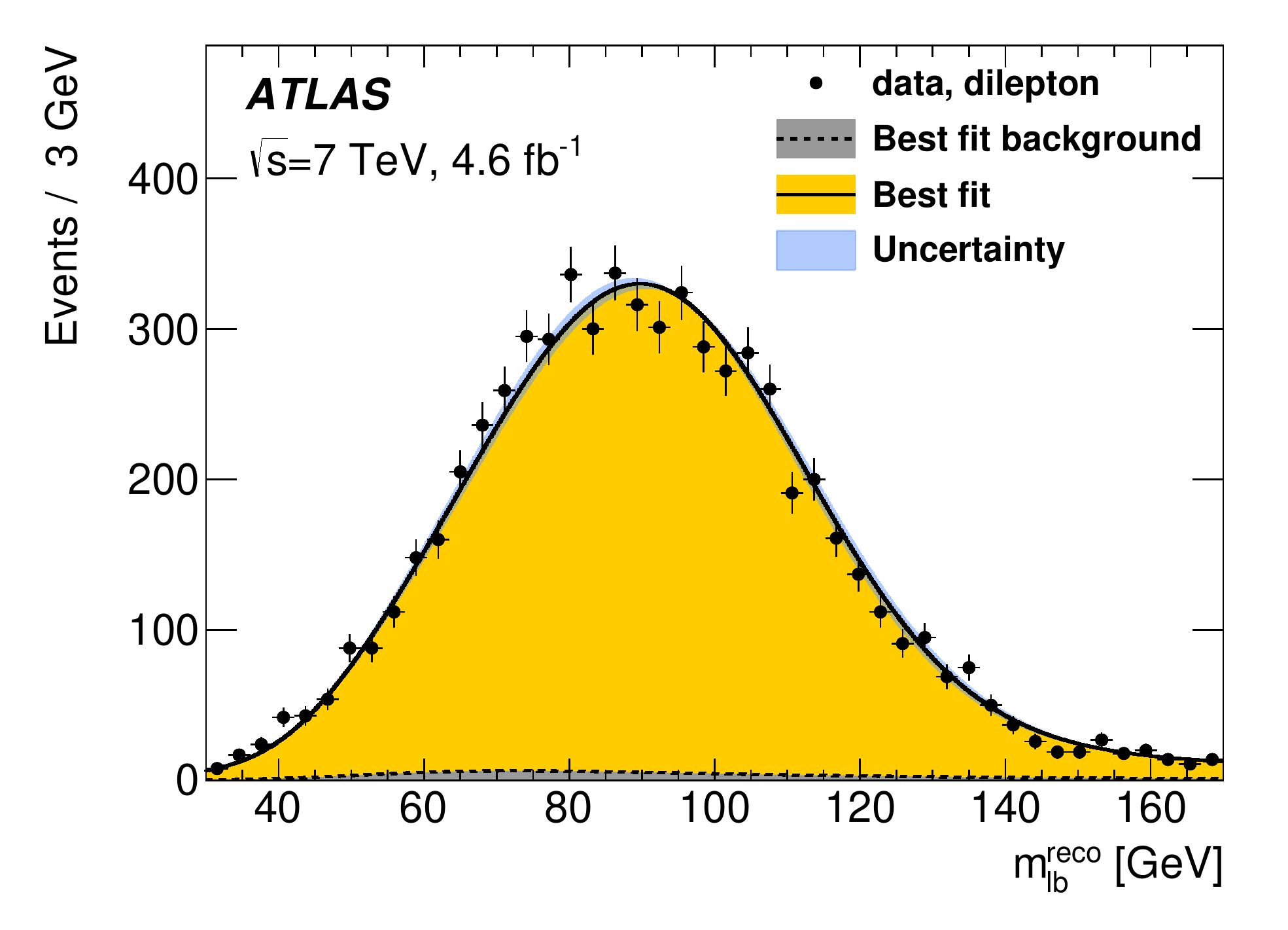}}
\caption{The fitted distributions in the data, showing (a) \mtr, (b) \mWr, (c) \rlbr,
  and (d) \mlbr.
  The fitted probability density functions for the background alone and for
  signal-plus-background are also shown.
  The uncertainty bands indicate the total uncertainty on the
  signal-plus-background fit obtained from pseudo-experiments as explained in
  the text. Figures~(a-c) refer to the \ttbarlj\ analysis, figure (d) to
  the \ttbarll\ analysis.  \label{Fig:mjjbdata}}
\end{figure*}
%
%
\begin{figure*}[tbp!]
  \centering \subfloat[\mtlj\ versus \JSF, \ljets\ ]{
    \includegraphics[width=0.49\textwidth]{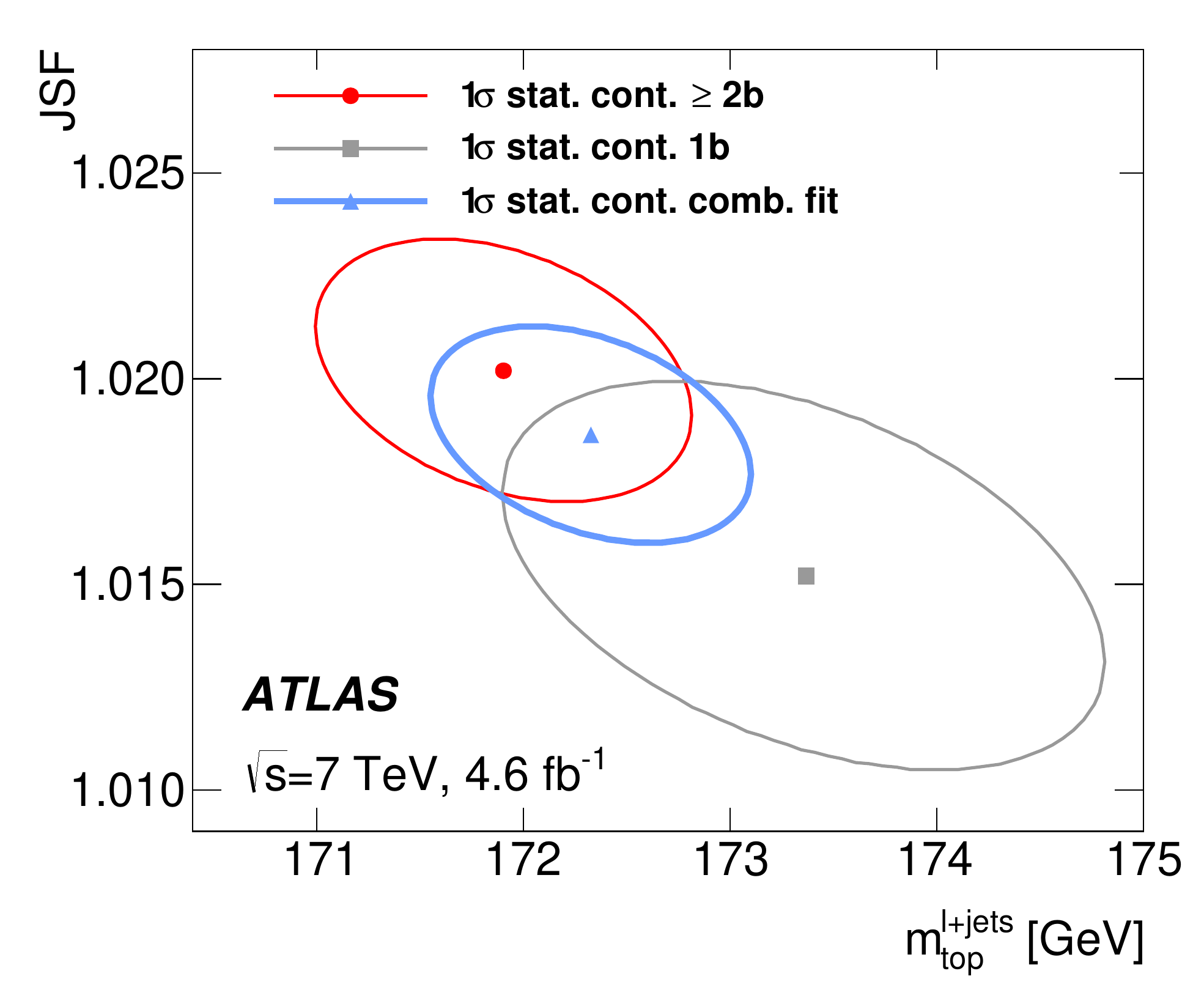}}
  \subfloat[\mt\ versus \bJSF, \ljets\ ]{
    \includegraphics[width=0.49\textwidth]{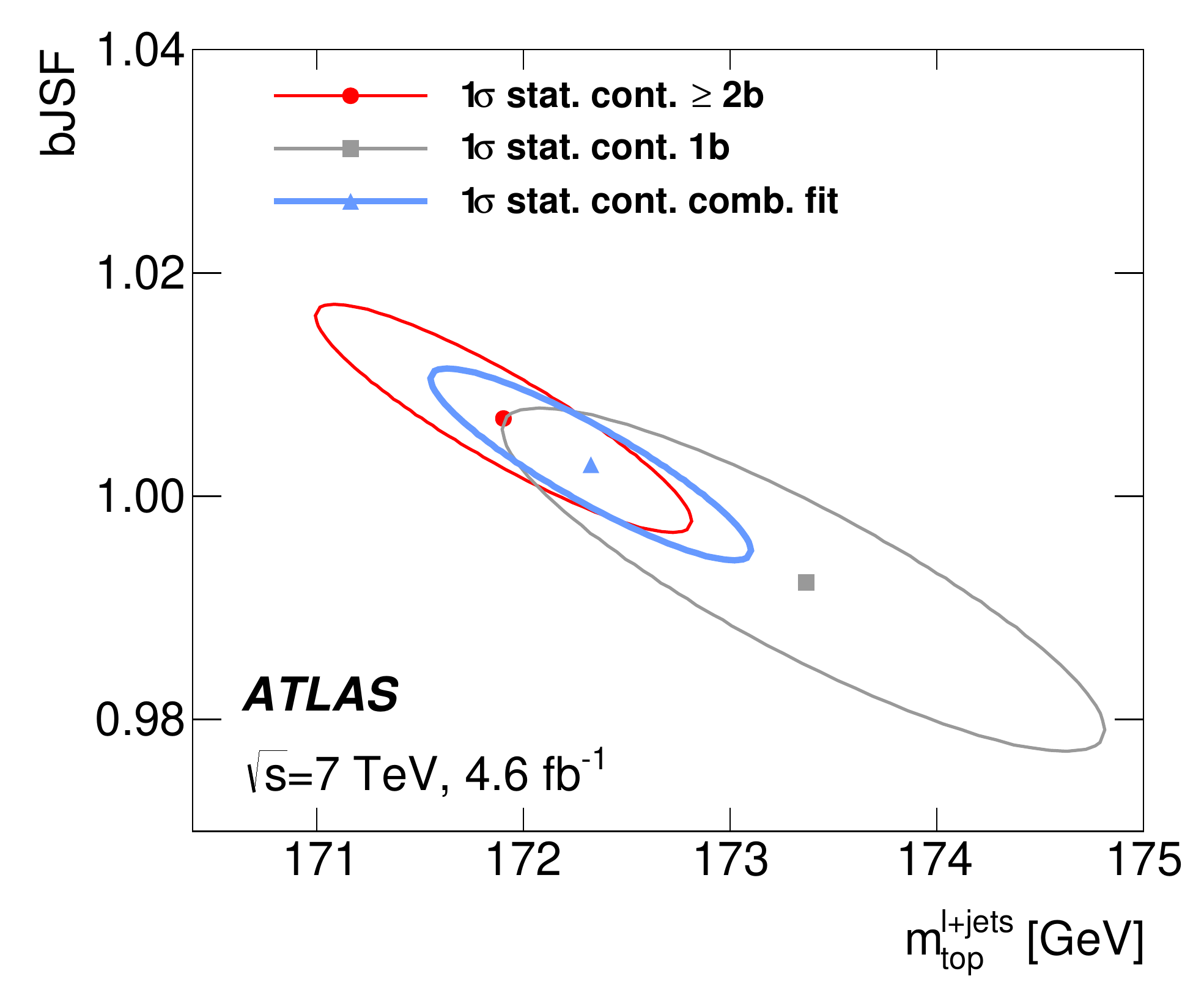}}
\hfill
  \subfloat[\JSF\ versus \bJSF, \ljets\ ]{
    \includegraphics[width=0.49\textwidth]{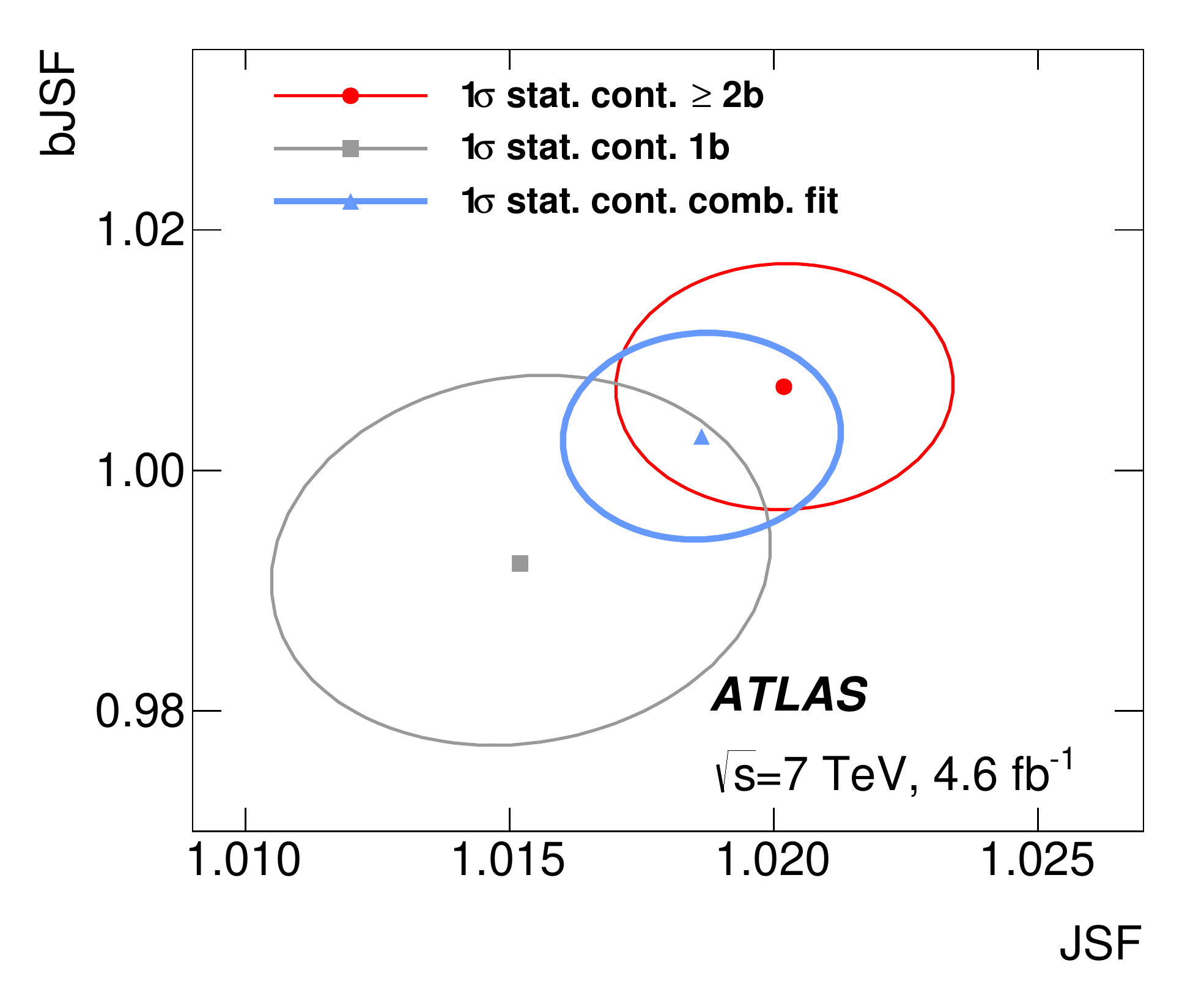}}
  \subfloat[\mtdl\ likelihood profile,  \dil\ ]{
    \includegraphics[width=0.49\textwidth]{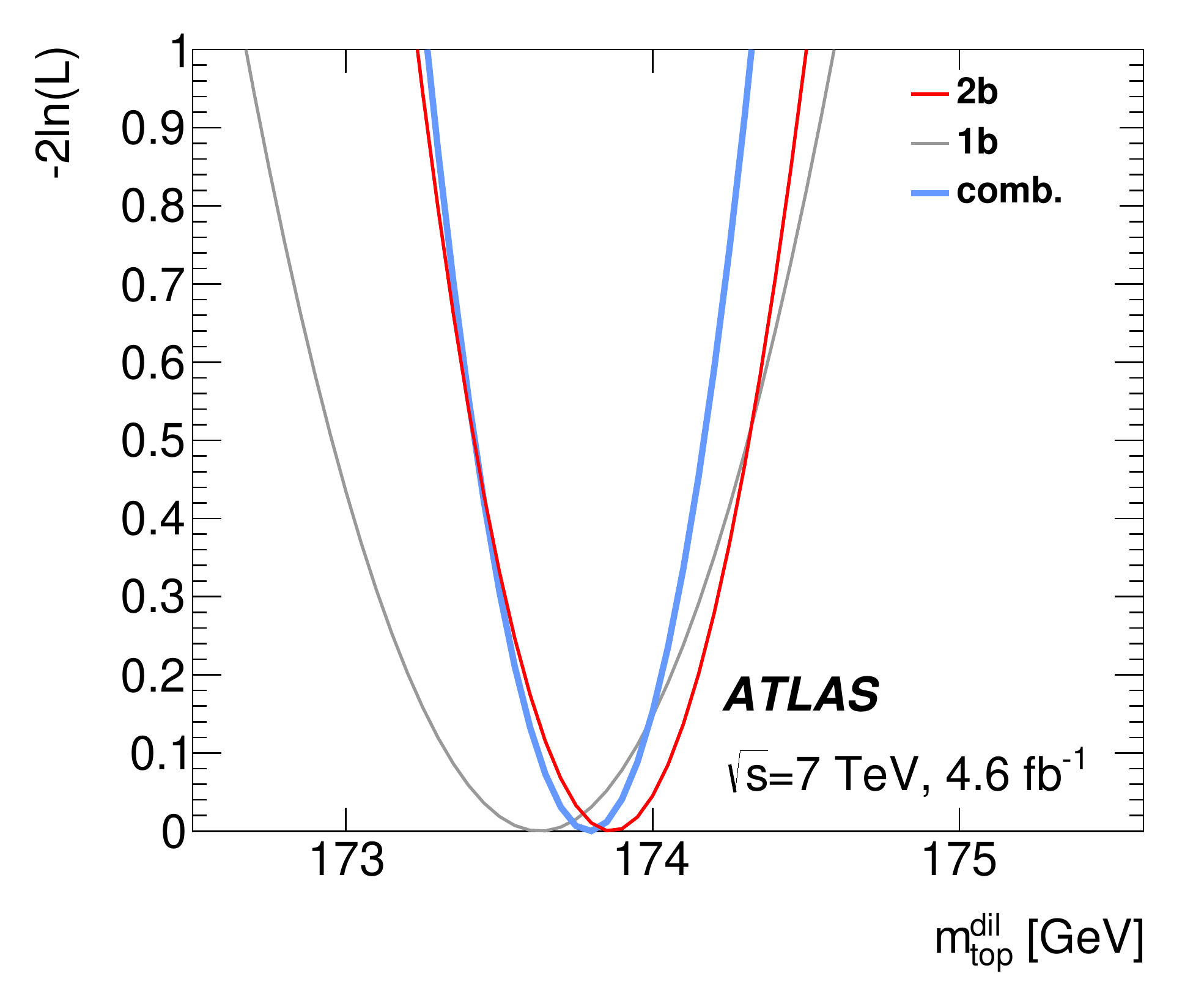}}
\caption{Likelihood contours showing the correlation determined in data 
  of the measured \mtlj\ to (a) the \JSF\ and (b) the \bJSF, and (c) the
  correlation of the two scales \JSF\ and \bJSF, within the \ttbarlj\ analysis.
  Figures~(a-c) show the results using the events with one \btagged\ jet only
  (grey ellipses), with at least two \btagged\ jets (red ellipses) and finally
  with all selected events, i.e.~the ones with at least one \btagged\ jet (blue
  ellipses). The ellipses correspond to the $\pm 1\sigma$ (statistical) uncertainties,
  including the statistical components from the \JSF\ and \bJSF\
  determination. While tracing the contours the additional parameters of the
  likelihood are fixed to their best fit values.
  Figure~(d) reports the likelihood profile as a function of \mtdl\ for the
  sample with one \btagged\ jet, the sample with two \btagged\ jets and the
  combined result.
  The colour coding is analogous to figures (a-c).
\label{Fig:contourdata}}
\end{figure*}

For the \ttbarlj\ analysis, the measured values of the three observables
(\mtlj, \JSF, and \bJSF), together with two-dimensional statistical uncertainty
contours ($\pm 1\sigma$), including the statistical components from the \JSF\
and \bJSF\ determination, are shown in Fig.~\ref{Fig:contourdata}(a--c).
Correspondingly, the likelihood profile as a function of \mtdl\ is reported in
Fig.~\ref{Fig:contourdata}(d), for the sample with one \btagged\ jet, the sample
with two \btagged\ jets and the combined \ttbarll\ result.
These results demonstrate the good agreement between the parameter values
measured in the samples with different \btagged\ jet multiplicities.

\section{Uncertainties affecting the \mt\ determination}
\label{sec:syst}
Several sources of systematic uncertainty are considered.  Their effects on
the \ljets\ and \dil\ measurements are listed in Table~\ref{tab:results},
together with the result of the combination of the two channels discussed in
Sect.~\ref{sec:resultcomb}.
Each source of uncertainty considered is investigated, when possible, by varying
the relevant quantities by $\pm 1\sigma$~with respect to their default values.
Using the changed parameters, 500 pseudo-experiments are performed using events
drawn from the full simulated samples.
The difference of the average \mt\ computed from pseudo-experiments based on the
standard MC sample, and the varied sample under consideration, both evaluated
with the original template parameterisations, is used to determine the
corresponding uncertainty. Unless stated otherwise, the systematic uncertainties
arising from the different modelling sources are calculated as half of the
difference of the results of the upward and downward variations.
The systematic uncertainties for the measured \JSF and \bJSF\ in the
\ttbarlj\ final state are also estimated.
Following Ref.~\cite{Barlow:2002yb}, the actual observed difference is quoted as
the systematic uncertainty on the corresponding source, even if it is smaller
than its associated statistical precision. The latter is estimated taking into
account the statistical correlation of the MC samples used in the comparison.
The total uncertainty is calculated as the sum in quadrature of all individual
contributions, i.e.~neglecting possible correlations (small by construction).
The estimation of the uncertainties for the individual contributions is
described in the following.
%
\subsection{Statistics and method calibration}
\paragraph{Statistical components due to the jet energy scale factors}\mbox{}\\
 The statistical uncertainty quoted for the \ttbarlj\ analysis is made up of
 three parts: a purely statistical component on \mt\ and the contributions
 stemming from the simultaneous determination of the \JSF\ and \bJSF.
 The former is obtained from a one-dimensional template method exploiting only
 the \mtr\ observable (fixing the values of the \JSF\ and \bJSF\ to the results
 of the three-dimensional analysis).
 The contribution to the statistical uncertainty on the fitted parameters due to
 the simultaneous fit of \mt\ and \JSF, is estimated as the difference in
 quadrature of the statistical uncertainty of a two-dimensional (\mtr\ and \mWr,
 fixing the value of \bJSF) fit and the one-dimensional fit to the data
 described above.
 Analogously, the contribution of the statistical uncertainty due to the
 simultaneous fit of \bJSF\ together with \mt\ and \JSF, is defined as the
 difference in quadrature of the statistical uncertainties obtained in the
 three-dimensional and the two-dimensional (fixing \bJSF) fits to the data.
 This separation allows a direct comparison of the sensitivity of the \mt\
 estimator for any analysis, irrespective of the number of observables exploited
 by the fit. In addition, the sensitivity of the estimators for the global jet
 energy scales can be directly compared.
 These uncertainties can be treated as uncorrelated uncertainties in \mt\
 combinations.  Together with the systematic components of the residual jet
 energy scale uncertainty discussed in Sect.~\ref{subsec:detectormodelling}
 below, they directly replace the uncertainty on \mt\ from the jet energy scale
 variations present without the in situ determination.
\paragraph{Method calibration}\mbox{}\\
This uncertainty takes into account the effect of any bias introduced in the fit
by the presence of correlations among the observables (neglected in the fit for
the \ttbarlj\ analysis) as well as the impact of the limited size of the MC
samples (for both analyses).  This leads to a systematic uncertainty in the
template fit, which is reflected in the residual mass differences of the fitted
mass and the input mass for a given MC sample. The largest average difference
observed in the pseudo-experiments carried out varying the underlying top quark
mass, the \JSF\ and the \bJSF\ with respect to the respective input parameter,
is taken as the uncertainty from this source.
%
\subsection{\ttbar\ modelling}
\paragraph{Signal Monte Carlo generator}\mbox{}\\
The systematic uncertainty related to the choice of \ttbar\ signal generator
program is determined by comparing the results of pseudo-experiments performed
with either the
\Mcatnlo~\cite{FRI-0201,FRI-0301} samples or the \Powheg\
samples, both generated with $\mt=172.5$~\GeV\ and using the
\Herwig\ program to perform the hadronisation.
This choice is supported by the observation that these \Mcatnlo\ and
\Powheg\ samples exhibit very different jet multiplicities for the
\ttbarlj\ channel which bracket those observed in
data~\cite{PaperOfATLAS-CONF-2012-155}.
The full difference of the results averaged over all pseudo experiments is
quoted as the systematic uncertainty.

The impact of changing the factorisation and renormalisation scales
($\mu_{\mathrm{F/R}}$) in \Powheg\ was also checked. The resulting
\mt\ systematic uncertainties amount to $0.15 \pm 0.07~\GeV$
and $0.14\pm 0.05~\GeV$ for the \ttbarlj\ channel, and \ttbarll\ analysis
respectively. Within the quoted statistical uncertainties, the $\mu_{\mathrm{
F/R}}$ systematic uncertainties are consistent with those originating from the
comparison of \Mcatnlo\ and \Powheg, which are used here.
\paragraph{Hadronisation}\mbox{}\\
Signal samples for $\mt=172.5$~\GeV\ from the \Powheg\ event generator
are produced performing the parton showering and the hadronisation
with either \Pythia\ with the P2011C tune or \Herwig\ and \Jimmy\ with
the ATLAS AUET2 tune~\cite{ATL-PHYS-PUB-2011-008}.
The full difference of the results averaged over all pseudo experiments is
quoted as the systematic uncertainty.
\paragraph{Initial- and final-state QCD radiation}\mbox{}\\
Different amounts of initial- and final-state QCD radiation can alter the jet
energies and multiplicities of the events, introducing distortions into the
measured \mtr, \mWr, \rlbr\ and \mlbr\ distributions.
This effect is evaluated by performing pseudo-experiments using two dedicated
signal samples generated with \Acermc~\cite{SAMPLES-ACER} in combination
with \Pythia\ P2011C for hadronisation and parton showering.  In these samples
some \Pythia\ P2011C parameters that control the showering are varied in ranges
that are compatible with a study of additional jets in \ttbar\
events~\cite{ATL-2012-033}, and half the difference of these two extremes
is used as the systematic uncertainty.
\paragraph{Underlying event and colour reconnection}\mbox{}\\
These systematic uncertainties are estimated using samples simulated
with \Powheg-hvq and \Pythia.  The underlying-event uncertainty is obtained by
comparing a sample with the Perugia 2012 tune (P2012) to a sample with the
P2012 mpiHi tune~\cite{Skands}. The full difference in the fitted mass of the two
models is taken as the systematic uncertainty for this source. Similarly, the
colour reconnection systematic uncertainty is assigned as the difference in the
fitted parameters of samples obtained with the P2012 and P2012 loCR
tunes~\cite{Skands}.
The same matrix-element-level \Powheg-hvq events generated with the CT10 PDFs
are used for the three MC samples.
The P2012 mpiHi tune is a variation of the P2012 tune with more semi-hard
multiple parton interactions.  The colour reconnection parameters were kept
fixed to the P2012 tune values. Compared to the standard P2012 tune the P2012
loCR tune leads to significantly less activity in the transverse region with
respect to the leading charged-particle as measured in
Ref.~\cite{ATL-PHYS-PUB-2011-009}. In addition to assessing the effect of colour
reconnection, this tune is therefore also used to estimate the systematic
uncertainty associated with the particle spectra in the underlying event.
\paragraph{Parton distribution functions}\mbox{}\\
The signal samples are generated using the CT10 PDFs. These PDFs, obtained from
experimental data, have an uncertainty that is reflected in 26~pairs of possible
PDF variations provided by the CTEQ group.
To evaluate the impact of the PDF uncertainty on the \ttbar\ signal templates,
the events, from a sample generated using \Mcatnlo\ with
\Herwig\ fragmentation, are re-weighted with the corresponding ratio
of PDFs, and 26~pairs of signal templates are constructed, one pair per PDF
uncertainty. For each pair, the average measured \mt\ is obtained from 500
pseudo-experiments each for the upward and downward variations of the PDF
uncertainty. The corresponding uncertainty is obtained as half the difference of
the two values. From those the CT10 contribution is calculated as the sum in
quadrature of the 26~uncertainties and amounts to 0.13~\GeV\ and 0.10~\GeV\ for
the \ttbarlj\ and \ttbarll\ analysis respectively.

In addition, the signal \ttbar\ samples are re-weighted to match the
central PDFs for either the MSTW2008~\cite{MAR-0901} or the
NNPDF23~\cite{Ball:2012cx} PDFs. The corresponding differences, taken as
uncertainties, are 0.03~\GeV\ and 0.21~\GeV\ for the \ttbarlj\ analysis, and
0.01~\GeV\ and 0.01~\GeV\ for the \ttbarll\ analysis.
The final PDF systematic uncertainty is the sum in quadrature of the three
contributions discussed above.
%
\subsection{Modelling of non-\ttbar\ processes}
The uncertainty in the modelling of non-\ttbar\ processes is taken into account
by varying the normalisation and the shape of the distributions of several
contributions.

The uncertainty on the \Wj\ background determined from
data~\cite{CERN-PH-EP-2012-015} is dominated by the uncertainty on the
heavy-flavour content of these events and amounts to \sysWnorm\ of the overall
normalisation.  The same normalisation uncertainty is assigned to the \Zj\
background normalisation.
Uncertainties related to the \Wj\ background shape are also considered. These
stem from the variation of the heavy-flavour composition of the samples and from
re-weightings of the distributions to match the predictions of \Alpgen. For the
re-weighting, parameters are varied which affect the functional form of the
factorisation and renormalisation scales, and the threshold for the matching
scale used to connect the matrix-element calculation to the parton shower.

The estimate of the background from NP/fake leptons determined from data is
varied by \uncqnorm\ to account for the uncertainty of this background
source~\cite{Aad:2010ey}.  Uncertainties affecting the shape of this background
are also included. For the NP/fake-electron background, the effects on the shape
arising from the efficiency uncertainties for real and fake electrons are
evaluated and added in quadrature. For the NP/fake-muon background, two different
matrix methods were used and averaged: their difference is taken as the
systematic uncertainty.

In addition, the impact of changing the normalisation of the single top quark
processes according to the uncertainty on the corresponding theoretical cross
sections is considered. This yields a negligible systematic uncertainty in both
the \ttbarlj\ and \ttbarll\ analyses.
%
\subsection{Detector modelling} \label{subsec:detectormodelling}
\paragraph{Jet energy scale}\mbox{}\\
The JES is derived using information from test-beam data, LHC collision data,
and simulation.
The relative JES uncertainty varies from about \uncjesglobmin\ to
\uncjesglobmax\ depending on jet \pt\ and $\eta$ as given in
Ref.~\cite{CERN-PH-EP-2013-222}.
Since the estimation of the jet energy scale involves a number of steps, the JES
uncertainty has various components originating from the calibration method, the
calorimeter response, the detector simulation, and the specific choice of
parameters in the physics model employed in the MC event generator.
The total uncertainty is expressed in terms of 21 \pt- and $\eta$-dependent
components which are considered uncorrelated~\cite{CERN-PH-EP-2013-222}.
The uncertainties for the individual components and their sum are given in
Table~\ref{tab:jesresults} in Appendix~\ref{app:JES}.
Despite the simultaneous fit of \mt, \JSF\ and \bJSF\ in the \ttbarlj\ channel
there is a non-negligible residual JES uncertainty.
This is introduced by the variation of the jet energy scale corrections and
their uncertainties with jet kinematics, which cannot be fully captured by
global scale factors (\JSF, \bJSF).
However the overall JES uncertainty is a factor of two smaller than in a \oned
exploiting only templates of \mtr.
In the \ttbarll\ channel, the contribution from the JES uncertainty constitutes
the main component of systematic uncertainty on \mt.
\paragraph{\bjet\ energy scale}\mbox{}\\
This uncertainty is uncorrelated with the JES uncertainty and accounts for the
remaining differences of \bjets\ and light-jets after the global JES was
determined.
For this, an extra uncertainty ranging from \uncbjesmin\ to
\uncbjesmax\ and depending on jet \pt\ and $\eta$ is assigned to
\bjets, due to differences between jets containing \bhadrons\ and the
inclusive jet sample~\cite{CERN-PH-EP-2013-222}.
This additional systematic uncertainty was obtained from MC simulation and was
verified using \btagged\ jets in data.
The validation of the \bjet\ energy scale uncertainty is based on the comparison
of the jet transverse momentum as measured in the calorimeter to the total
transverse momentum of charged-particles associated with the jet. These
transverse momenta are evaluated in the data and in MC simulated events for all
jets and for \bjets~\cite{CERN-PH-EP-2013-222}. In addition, a validation
using \ttbarlj\ events was performed.
Effects stemming from \bquark\ fragmentation, hadronisation and underlying soft
radiation were studied using different MC event generation
models~\cite{CERN-PH-EP-2013-222}.
Thanks to the simultaneous fit to \rlbr\ together with \mWr\ and \mtr,
the \ttbarlj\ \threed\ method mitigates the impact of this uncertainty, and
reduces it to 0.06~\GeV, instead of 0.88~\GeV\ in a
\twod\ method (exploiting two-dimensional templates of \mtr\ and \mWr,
as in Ref.~\cite{ATL-2012-Mass}), albeit at the cost of an additional
statistical component of 0.67~\GeV.
In the \ttbarll\ channel, the contribution from the bJES uncertainty represents
the second largest component of systematic uncertainty on \mt.
\paragraph{Jet energy resolution}\mbox{}\\
To assess the impact of this uncertainty, before performing the event selection,
the energy of each reconstructed jet in the simulation is smeared by a Gaussian
function such that the width of the resulting Gaussian distribution corresponds
to the one including the uncertainty on the jet energy
resolution~\cite{ATLAS-JER-2010}.
The fit is performed using smeared jets and the deviation from the central
result is assigned as a systematic uncertainty.
\paragraph{Jet reconstruction efficiency}\mbox{}\\
The jet reconstruction efficiency for data and the MC simulation is found to be
in agreement with an accuracy of better than \uncjeff~\cite{Aad:1409965}.
To account for the residual uncertainties, 2\% of jets with $\pt< 30$~\GeV\ are
randomly removed from MC simulated events.
The event selection and the fit are repeated on the changed sample. The changes
in the fitted parameters relative to the nominal MC sample are assigned as
systematic uncertainty.
\paragraph{Jet vertex fraction}\mbox{}\\
Residual differences between data and MC in the description of the fraction of
the jet momentum associated with tracks from the primary vertex (used to
suppress pile-up interactions) is corrected by applying scale factors. These
scale factors, varied according to their uncertainty, are applied to MC
simulation events as a function of the jet $\pT$. The resulting variation in the
measured top quark mass in the \ttbarlj\ analysis is 10~\MeV, while it is
negligible for the
\ttbarll\ analysis.
\paragraph{$b$-tagging efficiency and mistag rate}\mbox{}\\
To account for potential mismodelling of the \btag\ efficiency and the mistag
rate, \btag\ scale factors, together with their uncertainties, are derived per
jet~\cite{ATLAS-CONF-2012-040,ATLAS-CONF-2012-043,ATLAS-CONF-2014-004,ATLAS-CONF-2012-097}.
They are applied to the MC events and depend on the jet \pt\ and \eta\ and the
underlying quark flavour. In this analysis these correction factors are obtained
from dijet~\cite{ATLAS-CONF-2012-043} and \ttbarll\ events.
The same \btag\ calibrations are applied to both the \ljets\ and \dil\
final states.  The \ttbar-based calibrations are obtained using the
methodology described in Ref.~\cite{ATLAS-CONF-2014-004}, applied to
the 7~\TeV\ data. The statistical correlation stemming from the use of
partially overlapping data sets for the \ttbarll\ \mt\ analysis and
the \btag\ calibration is estimated to be negligible. 
The correlation of those systematic uncertainties that are in common for
the \btag\ calibration and the present analyses is taken into account.
Similarly to the JES uncertainty, the uncertainty on the correction factors for
the \btag\ efficiency is separated into ten uncorrelated components. The
systematic uncertainty is assessed by changing the correction factor central
values by $\pm1\sigma$ for each component, and performing the fit. The final
uncertainty due to the \btag\ efficiency is calculated as the sum in quadrature
of all contributions.
A similar procedure is applied for the mistag rates for \cjets, albeit using
four separate components.
In addition, the correction factors and mistag rates for light-jets are varied
within their uncertainty, and the corresponding shifts in the measured
quantities are summed in quadrature.
The size of the \btag\ systematic uncertainty of 0.50~\GeV\ observed in
the \ttbarlj\ analysis is mostly driven by the induced change in shape of
the \rlbr\ distribution.
%
\begin{table*}[tbp!]
\small
\begin{center}
\begin{tabular}{|l||r|r|r||r||r|r|}\cline{2-7}
\multicolumn{1}{c|}{}  & \multicolumn{3}{c||}{\ttbarlj} & \ttbarll & \multicolumn{2}{c|} {Combination}\\\cline{2-7}
\multicolumn{1}{c|}{}  &   \mtlj\ [\GeV] &   \JSF\ & \bJSF\   &  \mtdl\ [\GeV] & \mtcb\ [\GeV] & $\rho$  \\\hline
 Results          &   172.33         & 1.019          & 1.003 & 173.79  & 172.99  &\\ \hline
       Statistics &     0.75         & 0.003          & 0.008
       &    0.54 & 0.48 & 0                     \\ 
\multicolumn{1}{|l||}{\it $~~$-- Stat. comp. (\mt)}   &     {\it 0.23}    &    {\it n/a   } &     {\it n/a  }  &    {\it  0.54}& &\\ 
\multicolumn{1}{|l||}{\it $~~$-- Stat. comp. (\JSF)}  &     {\it 0.25}    &    {\it 0.003 } &     {\it n/a  }  &    {\it  n/a }& &\\
\multicolumn{1}{|l||}{\it $~~$-- Stat. comp. (\bJSF)} &     {\it 0.67}    &    {\it 0.000 } &     {\it 0.008}    &  {\it    n/a} & & \\
            Method &  0.11 $\pm$ 0.10 & 0.001   & 0.001  & 0.09 $\pm$
         0.07 & 0.07 & $ 0$   \\  \hline
         Signal MC &  0.22 $\pm$ 0.21 & 0.004   & 0.002  & 0.26 $\pm$ 0.16 & 0.24 & $+1.00$   \\
     Hadronisation &  0.18 $\pm$ 0.12 & 0.007   & 0.013  & 0.53 $\pm$ 0.09 & 0.34 & $+1.00$   \\
           ISR/FSR &  0.32 $\pm$ 0.06 & 0.017   & 0.007  & 0.47 $\pm$ 0.05 & 0.04 & $-1.00$   \\
  Underlying event &  0.15 $\pm$ 0.07 & 0.001   & 0.003  & 0.05 $\pm$ 0.05 & 0.06 & $-1.00$   \\
Colour reconnection&  0.11 $\pm$ 0.07 & 0.001   & 0.002  & 0.14 $\pm$ 0.05 & 0.01 & $-1.00$   \\
               PDF &  0.25 $\pm$ 0.00 & 0.001   & 0.002  & 0.11 $\pm$ 0.00 & 0.17 & $+0.57$   \\ \hline
   $W/Z$+jets norm &  0.02 $\pm$ 0.00 & 0.000   & 0.000  & 0.01 $\pm$ 0.00 & 0.02 & $+1.00$   \\
  $W/Z$+jets shape &  0.29 $\pm$ 0.00 & 0.000   & 0.004  & 0.00 $\pm$ 0.00 & 0.16 &  0   \\
NP/fake-lepton norm. &  0.10 $\pm$ 0.00 & 0.000   & 0.001  & 0.04 $\pm$ 0.00 & 0.07 & $+1.00$   \\
NP/fake-lepton shape &  0.05 $\pm$ 0.00 & 0.000   & 0.001  & 0.01 $\pm$ 0.00 & 0.03 & $+0.23$   \\ \hline
  Jet energy scale &  0.58 $\pm$ 0.11 & 0.018   & 0.009  & 0.75 $\pm$ 0.08 & 0.41 & $-0.23$   \\
$b$-jet energy scale &  0.06 $\pm$ 0.03 & 0.000   & 0.010  & 0.68 $\pm$ 0.02 & 0.34 & $+1.00$   \\
    Jet resolution &  0.22 $\pm$ 0.11 & 0.007   & 0.001  & 0.19 $\pm$ 0.04 & 0.03 & $-1.00$   \\
    Jet efficiency &  0.12 $\pm$ 0.00 & 0.000   & 0.002  & 0.07 $\pm$ 0.00 & 0.10 & $+1.00$   \\
Jet vertex fraction&  0.01 $\pm$ 0.00 & 0.000   & 0.000  & 0.00 $\pm$ 0.00 & 0.00 & $-1.00$   \\
             \btag &  0.50 $\pm$ 0.00 & 0.001   & 0.007  & 0.07 $\pm$ 0.00 & 0.25 & $-0.77$   \\
              \met &  0.15 $\pm$ 0.04 & 0.000   & 0.001  & 0.04 $\pm$ 0.03 & 0.08 & $-0.15$   \\
           Leptons &  0.04 $\pm$ 0.00 & 0.001   & 0.001  & 0.13 $\pm$ 0.00 & 0.05 & $-0.34$   \\

           Pile-up &  0.02 $\pm$ 0.01 & 0.000   & 0.000  & 0.01 $\pm$ 0.00 & 0.01 & $ 0$   \\\hline
             Total &  1.27 $\pm$ 0.33 & 0.027   & 0.024  & 1.41 $\pm$ 0.24 & 0.91 & $-0.07$\\\hline
\end{tabular}
\end{center}
\caption{The measured values of \mt\ and the contributions of various 
  sources to the uncertainty in the \ttbarlj\ and the \ttbarll\ analyses. The
  corresponding uncertainties in the measured values of the \JSF and \bJSF\ are
  also shown for the \ttbarlj\ analysis. The statistical uncertainties
  associated with these values are typically 0.001 or smaller.
  The result of the \mt\ combination is shown in the rightmost columns, together
  with the correlation ($\rho$) within each uncertainty group as described in
  Sect.~\protect\ref{sec:resultcomb}. The symbol n/a stands for not
  applicable. Values quoted as 0.00 are smaller than 0.005.
  Finally, the last line refers to the sum in quadrature of the statistical and
  systematic uncertainty components. \label{tab:results}}
\end{table*}
%
\paragraph{Lepton momentum  and missing transverse momentum }\mbox{}\\
The lepton momentum and the \met\ are used in the event selection and
reconstruction.
For the leptons, the momentum scale, resolution and identification efficiency
are measured using high-purity $Z\to \ell\ell$ data~\cite{CERN-PH-EP-2014-040,
CERN-PH-EP-2014-151}.
The uncertainty due to any possible miscalibration is propagated to the analyses
by changing the measured reconstruction efficiency, lepton $\pt$, and the
corresponding resolution, within uncertainties.

The uncertainties from the energy scale and resolution corrections for leptons
and jets are propagated to the \met. The systematic uncertainty related to
the \met\ accounts for uncertainties in the energies of calorimeter cells not
associated with the reconstructed objects, and from cells associated with
low-\pt\ jets (7 GeV< \pt < 20 GeV), as well as for the dependence of their
energy on the number of pile-up interactions~\cite{ATLAS-MET-NEW}.
\paragraph{Pile-up}\mbox{}\\
The residual systematic uncertainty due to pile-up was assessed by determining
the dependence of the fitted top quark mass on the amount of pile-up activity,
combined with uncertainties in modelling the amount of pile-up in the sample.
%
\subsection{Summary}
The resulting sizes of all uncertainties and their sum in quadrature are given
in Table~\ref{tab:results}.  The total uncertainties on \mtlj,
\JSF, \bJSF\ and \mtdl, amount to \mtunc~\GeV, \jsfunc, \bjsfunc\ and
\mtdlunc~\GeV, respectively. Within uncertainties, the fitted values
of \JSF\ and \bJSF\ are consistent with unity.
%

\section{Combination of the \mt\ results}
\label{sec:resultcomb}
 The results of the \ttbarlj\ and \ttbarll\ analyses listed in
 Table~\ref{tab:results} are combined using the Best Linear Unbiased Estimate
 (BLUE) method~\cite{BLUE1,BLUE2}, implemented as described in
 Refs.~\cite{BLUEcpp,BLUERN}.
 The BLUE method determines the coefficients (weights) to be used in a linear
 combination of the input measurements by minimising the total uncertainty of
 the combined result. In the algorithm, both the statistical and systematic
 uncertainties, and the correlations ($\rho$) of the measurements, are taken
 into account, while assuming that all uncertainties are distributed according
 to Gaussian probability density functions.
%
\subsection{Correlation of the \ttbarlj\ and \ttbarll\ measurements}
\label{sec:sgncorr}
 To perform the combination, for each source of systematic uncertainty, the
 uncertainties as well as the correlation of the measurements of \mt\ were
 evaluated.

 The measurements are taken as uncorrelated for the statistical, the
 method calibration and the pile-up uncertainties.
 For the remaining uncertainty components there are two possible
 situations. Either the measurements are fully correlated, $\rho=+1$, i.e.~a
 simultaneous upward variation of the systematic uncertainty results in a
 positive (or negative) shift of \mt\ for both measurements, or fully
 anti-correlated, $\rho=-1$. In the latter case one measurement exhibits a
 positive shift and the other a negative one.

 Figure~\ref{fig:LJDLsystCorrTG}(a) shows the two dimensional distribution of
 the systematic uncertainties, denoted by \dmtlj\ and \dmtdl, obtained in
 the \ljets\ and \dil\ analyses for all components of the sources of systematic
 uncertainty for which the measurements are correlated.
 The points show the estimated size of the uncertainties, and the error bars
 represent the statistical uncertainties on the estimates.
 Some uncertainty sources in Table~\ref{tab:results}, such as the uncertainty
 related to the choice of MC generator for signal events, contain only a single
 component. For these type of sources, the correlation is either $\rho=+1$
 (red points) or $\rho=-1$ (blue points).
 The size of the uncertainty bars in Fig.~\ref{fig:LJDLsystCorrTG}(a) indicates
 that the distinction between $\rho=+1$ and $\rho=-1$ can be unambiguously made
 for all components that significantly contribute to the systematic uncertainty
 on \mt.
%
\begin{figure*}[t!]
\centering
\subfloat[3d \ljets\ vs \dil\ ]{
  \includegraphics[width=0.33\textwidth]{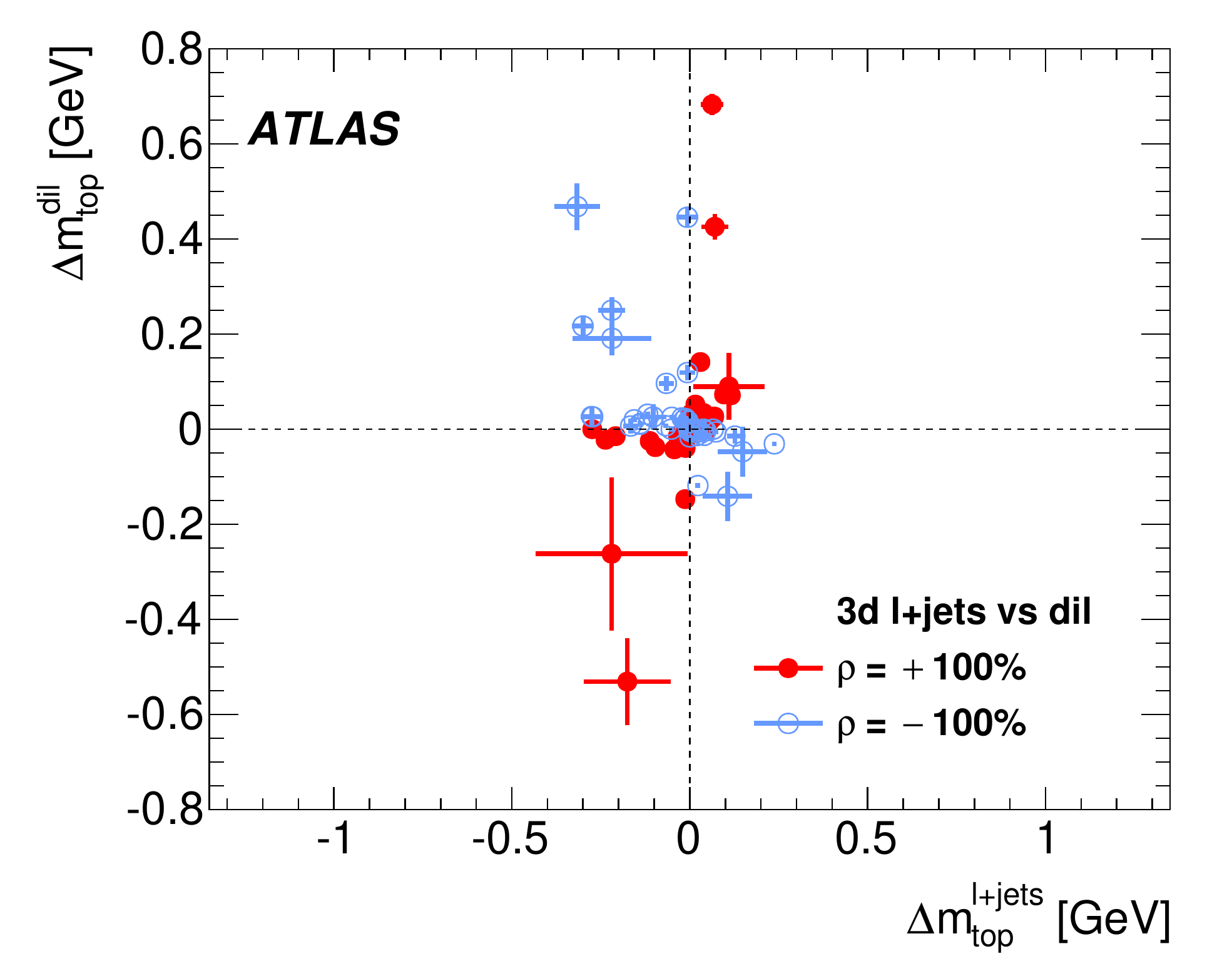}}
\subfloat[2d \ljets\ vs \dil\ ]{
  \includegraphics[width=0.33\textwidth]{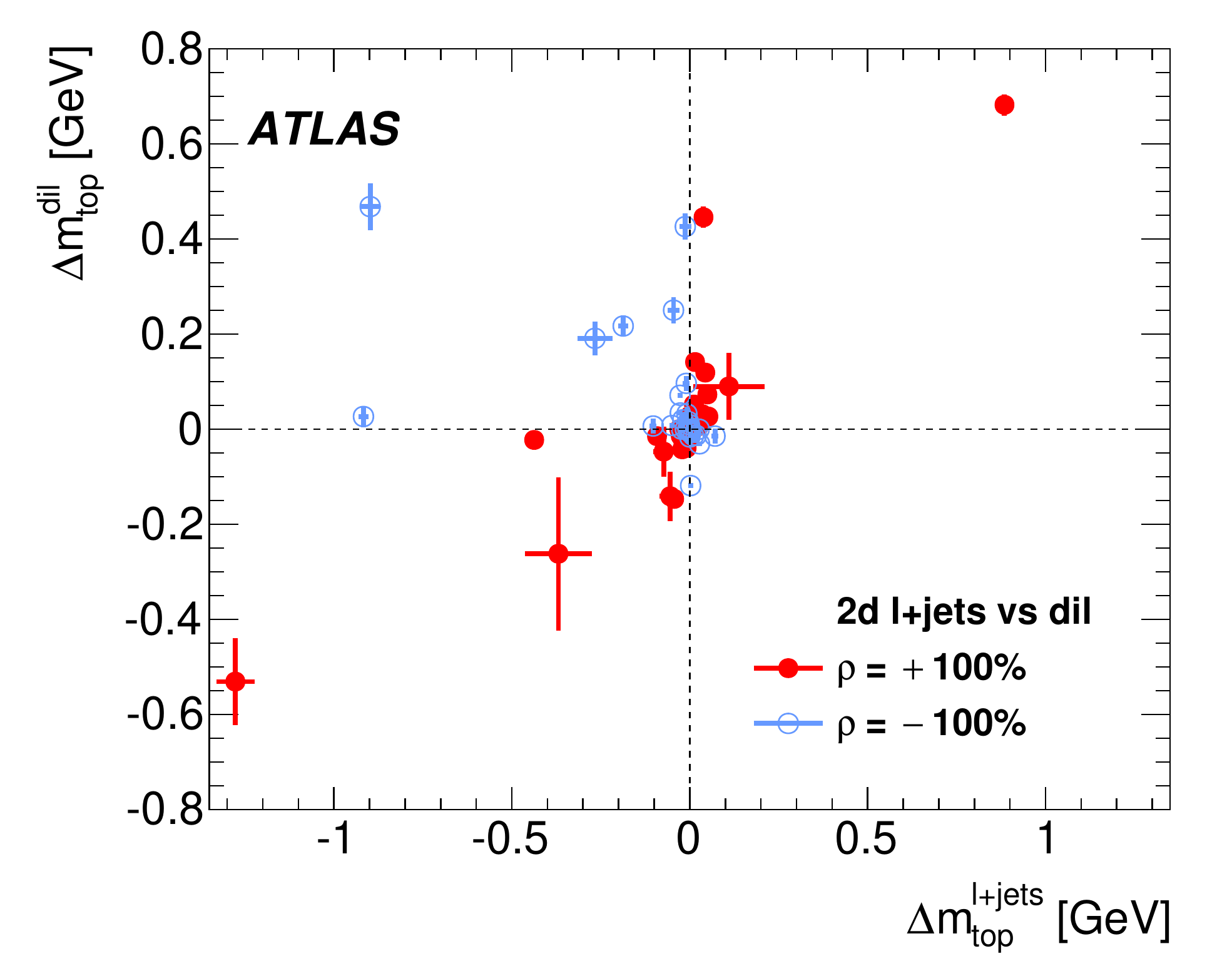}}
\subfloat[1d \ljets\ vs \dil\ ]{
  \includegraphics[width=0.33\textwidth]{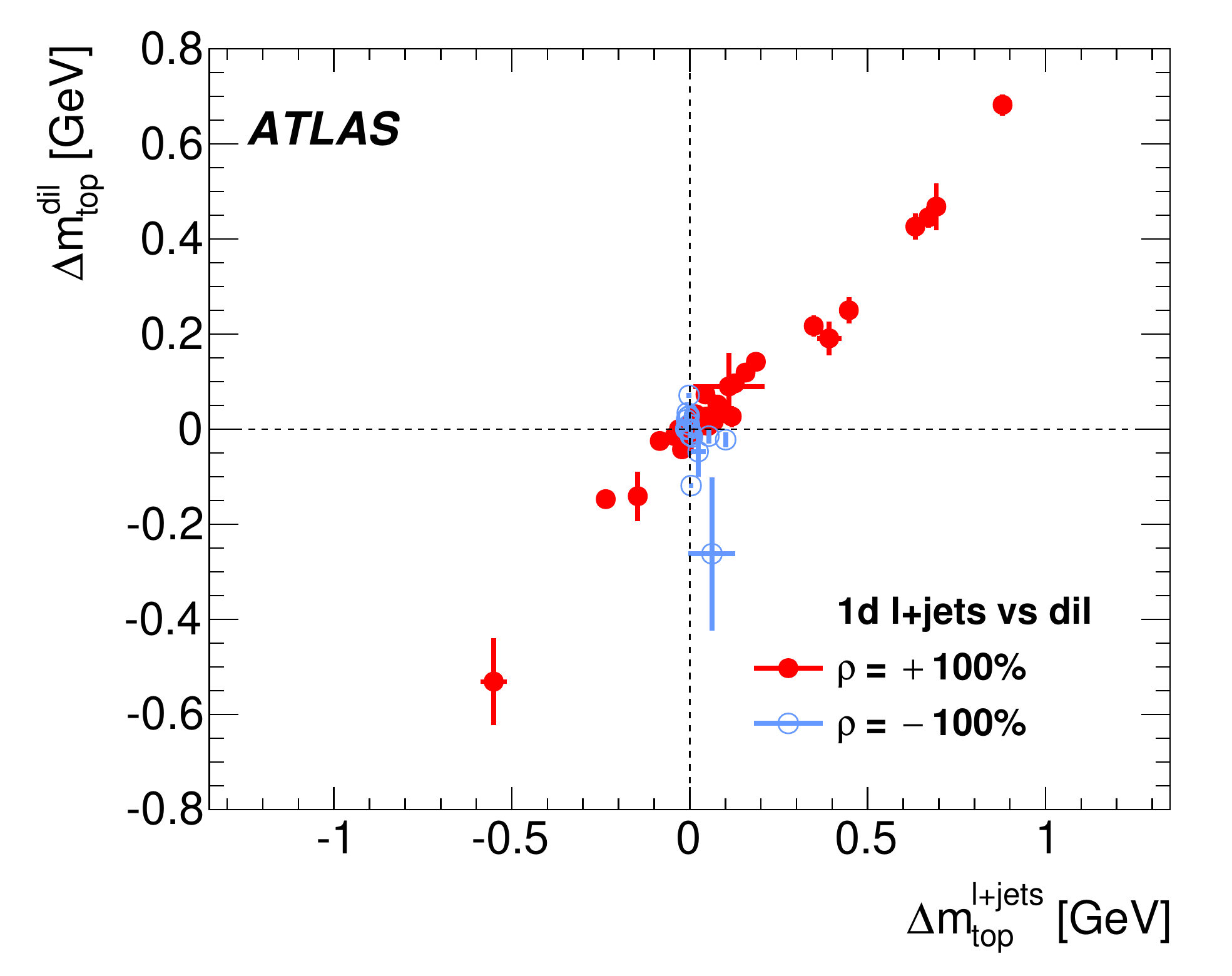}}
\caption{The systematic uncertainties of \mt\ in the \ljets\ analysis versus
  those of the \dil\ analysis.
  Figures~(a--c) refer to the results evaluated for the \threed~(3d), \twod~(2d)
  and \oned~(1d).
  The points show the estimated systematic uncertainties on \mt\ for the two
  analyses, and the uncertainty bars reflect the corresponding statistical
  uncertainties. The different colours reflect the different correlations
  described in Sect.~\protect\ref{sec:sgncorr}.  \label{fig:LJDLsystCorrTG}}
\end{figure*}
%
 For uncertainty sources that contain multiple components such as the JES
 uncertainty described in Appendix~\ref{app:JES}, the correlations given in
 Table~\ref{tab:results} differ from $\rho=\pm 1$. For these cases the
 correlation is obtained by adding the corresponding covariance terms of the
 components and dividing by the respective total uncertainties of the source.

 For each systematic uncertainty, the size of \dmtlj\ and \dmtdl, and the
 correlation of the measurements depend on the details of the analyses.
 This can be seen from Figs.~\ref{fig:LJDLsystCorrTG}(b, c) where the same
 information as in Fig.~\ref{fig:LJDLsystCorrTG}(a) is shown, but for different
 implementations of the \ljets\ analysis, while leaving the \dil\ analysis
 unchanged.
 Figure~\ref{fig:LJDLsystCorrTG}(b) corresponds to a \twod, similar to
 Ref.~\cite{ATL-2012-Mass}, which is realised by fixing the \bJSF\ to unity.
 Finally, Fig.~\ref{fig:LJDLsystCorrTG}(c) shows the result of a \oned, in which
 the values of the \JSF\ and \bJSF\ are fixed to unity. For this
 implementation, as for the \dil analysis, only \mt\ is obtained from the fit to
 data.
 Compared to the \twod, the \threed\ reduces some sources of uncertainty
 on \mt. As an example, the rightmost red point in
 Fig.~\ref{fig:LJDLsystCorrTG}(b), which corresponds to the bJES uncertainty,
 lies close to the vertical line in Fig.~\ref{fig:LJDLsystCorrTG}(a), i.e.~for
 the \ljets\ analysis the impact of this source was considerably reduced by
 the \bJSF\ determination from data.
 The change in the correlations of the measurements for specific sources of
 uncertainty, caused by a variation of the analysis strategy, is apparent from
 Fig.~\ref{fig:LJDLsystCorrTG}(c), where for both analyses only \mt\ is
 obtained from the data.
 In this case the exploited observables are much more similar and consequently,
 the measurements of \mt\ are fully correlated for all sources of uncertainty
 that significantly contribute to the total uncertainty.
 This demonstrates that the \threed\ not only reduces the impact of some sources
 of uncertainty, mainly the JES and bJES uncertainties, but also makes the two
 measurements less correlated, thus increasing the gain in the combination of
 the two estimates of \mt.

 To best profit from the combination of the two measurements, their correlation
 should be as small as possible, see Ref.~\cite{BLUERN}. Consequently, the jet
 energy scale factors measured in the \ljets\ analysis have not been propagated
 to the dilepton analysis, as was first done in Ref.~\cite{CDFsimul2009}.
 Transferring the scales would require adding an additional systematic
 uncertainty to the \dil\ analysis to account for the different jet energy scale
 factors caused by different kinematical selections and jet topologies of the
 two analyses. The two final states contain either two or four jets that have
 different distributions in jet \pT, and different amounts of final state QCD
 radiation. Most notably, this would also result in a large correlation of the
 measurements, similar to that observed for the one-dimensional analyses shown
 in Fig.~\ref{fig:LJDLsystCorrTG}(c).
 Consequently, the knowledge of \mt\ from the \ljets\ analysis would not
 significantly improve when including a \dil\ measurement obtained with
 transferred jet energy scales. For an example of such a situation see Table VI
 of Ref.~\cite{CDFsimul2009}.

 Using the correlations determined above, the combination of the \mt\ results of
 the \ttbarlj\ and \ttbarll\ analyses yields:

 $$\mtcb = 172.99 \pm 0.48~(\rm stat) \pm 0.78~(\rm syst) \GeV = 172.99 \pm
 0.91 \GeV.$$
 This value corresponds to a $28\%$ gain in precision with respect to the more
 precise \ljets\ measurement. The compatibility of the input measurements is
 very good, and corresponds to $0.75\sigma$ ($\mtlj-\mtdl = -1.47 \pm
 1.96~\GeV$).
 The BLUE weights of the results of the \ttbarlj\ and \ttbarll\ analyses are
 54.8\% and 45.2\%, respectively. The total correlation of the input
 measurements is $-7\%$ and the $\chi^2$ probability of the combination is
 45.5\%.
 The list of all uncertainties of the combined result, together with the
 correlation of the measurements for each group of uncertainties, is provided in
 Table~\ref{tab:results}. The current precision is mostly limited by systematic
 uncertainties related to the MC modelling of \ttbar\ events, and to the
 calibration of the jet energy scales.
%
\subsection{Stability of the results}
\label{sec:stabtests}
 The dependence of the combined result on the statistical uncertainties of the
 evaluated systematic uncertainties is investigated by performing one thousand
 BLUE combinations in which all input uncertainties are independently smeared
 using Gaussian functions centred at the expected values, and with a width
 corresponding to their statistical uncertainties.
 Using the smeared uncertainties, the correlations are re-evaluated for each
 pseudo-experiment. The combined \mt\ and its total uncertainty are distributed
 according to Gaussian functions of width 37~\MeV\ and 43~\MeV, respectively.
 Similarly, the BLUE combination weights and the total correlation are Gaussian
 distributed, with widths of 2.5$\%$ and 6.1$\%$, respectively.
 These effects are found to be negligible compared to the total uncertainty of
 the combined result. Consequently, no additional systematic uncertainty is
 assigned.


\section{Conclusion}
\label{sec:summary}
The top quark mass was measured via a three-dimensional template method in
the \ttbarlj\ final state, and using a one-dimensional template method in
the \ttbarll\ channel. Both analyses are based on $\sqrt{s}=7$~\TeV\
proton--proton collision ATLAS data from the 2011 LHC run corresponding to
an integrated luminosity of \atlumo~\ifb.
In the \ljets\ analysis, \mt\ is determined together with a global jet energy
scale factor (\JSF) and a residual $b$-to-light-jet energy scale factor (\bJSF).
The measured values are:
%
\begin{eqnarray*}
 \mtlj& = &  \mtval \pm \mtstaScales~\mathrm{(stat + \JSF + \bJSF)} 
            \pm \mtsys~\mathrm{(syst)}~\GeV, \\
 \JSF & = & \XZ{\jsfval}{\jsfsta}{\jsfsys},\\
 \bJSF & = & \XZ{\bjsfval}{\bjsfsta}{\bjsfsys}, \\
 \mtdl & = &  \mtdlval \pm \mtdlsta~\mathrm{(stat)} 
            \pm \mtdlsys~\mathrm{(syst)}~\GeV.
\end{eqnarray*}
%
These measurements are consistent with the ATLAS measurement in the fully
hadronic decay channel~\cite{ATLASaj2014}, and supersede the previous result
described in Ref.~\cite{ATL-2012-Mass}.

A combination of the \ttbarlj\ and \ttbarll\ results is performed using the BLUE
technique, exploiting the full uncertainty breakdown, and taking into account
the correlation of the measurements for all sources of the systematic
uncertainty.
The result is:

$$\mtcb = 172.99 \pm 0.48~(\rm stat) \pm 0.78~(\rm syst) \GeV = 172.99 \pm
0.91 \GeV.$$

This corresponds to a gain in precision with respect to the more precise \ljets\
measurement of $28\%$.
The total uncertainty of the combination corresponds to 0.91~\GeV\ and is
currently dominated by systematic uncertainties due to jet calibration and
modelling of the \ttbar\ events.


\section*{Acknowledgements}


We thank CERN for the very successful operation of the LHC, as well as the
support staff from our institutions without whom ATLAS could not be
operated efficiently.

We acknowledge the support of ANPCyT, Argentina; YerPhI, Armenia; ARC,
Australia; BMWFW and FWF, Austria; ANAS, Azerbaijan; SSTC, Belarus; CNPq and FAPESP,
Brazil; NSERC, NRC and CFI, Canada; CERN; CONICYT, Chile; CAS, MOST and NSFC,
China; COLCIENCIAS, Colombia; MSMT CR, MPO CR and VSC CR, Czech Republic;
DNRF, DNSRC and Lundbeck Foundation, Denmark; EPLANET, ERC and NSRF, European Union;
IN2P3-CNRS, CEA-DSM/IRFU, France; GNSF, Georgia; BMBF, DFG, HGF, MPG and AvH
Foundation, Germany; GSRT and NSRF, Greece; RGC, Hong Kong SAR, China; ISF, MINERVA, GIF, I-CORE and Benoziyo Center, Israel; INFN, Italy; MEXT and JSPS, Japan; CNRST, Morocco; FOM and NWO, Netherlands; BRF and RCN, Norway; MNiSW and NCN, Poland; GRICES and FCT, Portugal; MNE/IFA, Romania; MES of Russia and NRC KI, Russian Federation; JINR; MSTD,
Serbia; MSSR, Slovakia; ARRS and MIZ\v{S}, Slovenia; DST/NRF, South Africa;
MINECO, Spain; SRC and Wallenberg Foundation, Sweden; SER, SNSF and Cantons of
Bern and Geneva, Switzerland; NSC, Taiwan; TAEK, Turkey; STFC, the Royal
Society and Leverhulme Trust, United Kingdom; DOE and NSF, United States of
America.

The crucial computing support from all WLCG partners is acknowledged
gratefully, in particular from CERN and the ATLAS Tier-1 facilities at
TRIUMF (Canada), NDGF (Denmark, Norway, Sweden), CC-IN2P3 (France),
KIT/GridKA (Germany), INFN-CNAF (Italy), NL-T1 (Netherlands), PIC (Spain),
ASGC (Taiwan), RAL (UK) and BNL (USA) and in the Tier-2 facilities
worldwide.

\clearpage
\appendix
\part*{Appendix}
\addcontentsline{toc}{part}{Appendix}

\section{Jet energy scale uncertainty: detailed components}
\label{app:JES}
 The relative JES uncertainty varies from aboutabout \uncjesglobmin\ to
 \uncjesglobmax\ depending on jet properties as given in Section~13 of
 Ref.~\cite{CERN-PH-EP-2013-222}.
 These components correspond to the eigenvectors of the reduced covariance
 matrix for the JES uncertainties, as described in Section 13.3 of
 Ref.~\cite{CERN-PH-EP-2013-222}. 
 The initial sources of nuisance parameters (NP) originating from the in-situ
 determination of the JES are listed in Table~10 of
 Ref.~\cite{CERN-PH-EP-2013-222}. According to their nature, they are
 categorised into the classes: detector description, physics modelling,
 statistics and method, mixed detector and modelling. Finally, following
 Section~13.6 of Ref.~\cite{CERN-PH-EP-2013-222}, a reduction of the number of
 nuisance parameters is performed for each category giving various components.
 Their \pT\ dependences are given in Fig.~42 of Ref.~\cite{CERN-PH-EP-2013-222}.
 The total JES uncertainty is provided together with its 21 sub-components in
 Table~\ref{tab:jesresults}. Their separate effects on the fitted top quark mass
 are summed in quadrature to determine the total jet energy scale uncertainty
 given in Table~\ref{tab:results}.
 For further details about each component, see Ref.~\cite{CERN-PH-EP-2013-222}.
\renewcommand{\arraystretch}{0.98}
\begin{table*}[hp!]
\footnotesize
\begin{center}
\begin{tabular}{|l||r|r|r||r||r|r|}\cline{2-7}
\multicolumn{1}{c|}{}  & \multicolumn{3}{c||}{\ttbarlj} & \ttbarll & \multicolumn{2}{c|}{Combination} \\\cline{2-7}
\multicolumn{1}{c|}{}  &   $\Delta\mtlj$ [\GeV] &   $\Delta\JSF$ & $\Delta\bJSF$   &  $\Delta\mtdl$ [\GeV] & $\Delta\mtcb$[\GeV] & $\rho$  \\\hline
   Statistical (total)           &  $ 0.18 \pm 0.04$ & $ 0.003$ &  $ 0.001$ &  $ 0.16 \pm 0.03$ & 0.11  & $-0.25$ \\ \cline{2-7}       
-- Statistical NP1               &  $-0.17 \pm 0.02$ & $+0.002$ &  $+0.001$ &  $+0.01 \pm 0.02$ & 0.09  & $-1.00$ \\         
-- Statistical NP2               &  $+0.02 \pm 0.00$ & $+0.001$ &  $-0.000$ &  $+0.05 \pm 0.00$ & 0.03  & $+1.00$ \\         
-- Statistical NP3               &  $-0.01 \pm 0.02$ & $+0.001$ &  $+0.001$ &  $+0.12 \pm 0.02$ & 0.05  & $-1.00$ \\
-- $\eta$ inter-calibration (stat.) &  $-0.07 \pm 0.02$ & $+0.001$ &  $+0.001$ &  $+0.10 \pm 0.02$ & 0.01  & $-1.00$ \\ \hline         
   Modelling (total)             &  $ 0.31 \pm 0.06$ & $ 0.009$ &  $ 0.002$ &  $ 0.52 \pm 0.04$ & 0.26  & $-0.18$ \\ \cline{2-7} 
-- Modelling NP1                 &  $-0.30 \pm 0.03$ & $+0.006$ &  $+0.001$ &  $+0.22 \pm 0.02$ & 0.07  & $-1.00$ \\        
-- Modelling NP2                 &  $+0.03 \pm 0.02$ & $+0.002$ &  $-0.000$ &  $+0.14 \pm 0.02$ & 0.08  & $+1.00$ \\         
-- Modelling NP3                 &  $-0.01 \pm 0.02$ & $-0.002$ &  $-0.000$ &  $-0.15 \pm 0.02$ & 0.07  & $+1.00$ \\         
-- Modelling NP4                 &  $-0.01 \pm 0.00$ & $+0.000$ &  $+0.000$ &  $+0.02 \pm 0.00$ & 0.00  & $-1.00$ \\
-- $\eta$ inter-calibration (model) &  $+0.07 \pm 0.04$ & $+0.007$ &  $-0.001$ &  $+0.43 \pm 0.03$ & 0.23  & $+1.00$ \\ \hline         
   Detector (total)              &  $ 0.05 \pm 0.03$ & $ 0.007$ &  $ 0.001$ &  $ 0.45 \pm 0.04$ & 0.20  & $-0.19$ \\ \cline{2-7}        
-- Detector NP1                  &  $-0.01 \pm 0.03$ & $+0.007$ &  $+0.001$ &  $+0.45 \pm 0.02$ & 0.20  & $-1.00$ \\         
-- Detector NP2                  &  $-0.05 \pm 0.00$ & $+0.000$ &  $+0.001$ &  $+0.03 \pm 0.00$ & 0.02  & $-1.00$ \\ \hline        
   Mixed (total)                 &  $ 0.02 \pm 0.02$ & $ 0.001$ &  $ 0.001$ &  $ 0.03 \pm 0.02$ & 0.01  & $-0.80$ \\ \cline{2-7}        
-- Mixed NP1                     &  $-0.02 \pm 0.00$ & $+0.000$ &  $+0.001$ &  $+0.02 \pm 0.00$ & 0.00  & $-1.00$ \\         
-- Mixed NP2                     &  $+0.00 \pm 0.02$ & $+0.001$ &  $-0.000$ &  $+0.02 \pm 0.02$ & 0.01  & $+1.00$ \\   \hline      
   Single particle high-\pt      &  $+0.00 \pm 0.00$ & $+0.000$ &  $-0.000$ &  $+0.00 \pm 0.00$ & 0.00  & $+1.00$ \\  \hline
   Relative non-closure MC       &  $+0.00 \pm 0.02$ & $+0.001$ &  $-0.000$ &  $+0.03 \pm 0.02$ & 0.02  & $+1.00$  \\ \hline
   Pile-up (total)               &  $ 0.15 \pm 0.04$ & $ 0.001$ &  $ 0.002$ &  $ 0.04 \pm 0.03$ & 0.09  & $+0.03$ \\ \cline{2-7}
-- Pile-up: Offset($\mu$)        &  $-0.11 \pm 0.02$ & $-0.001$ &  $+0.001$ &  $-0.02 \pm 0.02$ & 0.07  & $+1.00$ \\         
-- Pile-up: Offset($n_{\rm vtx}$) &  $-0.10 \pm 0.04$ & $-0.000$ &  $+0.001$ &  $+0.03 \pm 0.03$ & 0.04  & $-1.00$ \\  \hline
   Flavour (total)               &  $ 0.36 \pm 0.04$ & $ 0.012$ &  $ 0.008$ &  $ 0.03 \pm 0.03$ & 0.20  & $-0.17$ \\ \cline{2-7}        
-- Flavour Composition           &  $-0.24 \pm 0.02$ & $+0.006$ &  $-0.002$ &  $-0.02 \pm 0.02$ & 0.14  & $+1.00$ \\         
-- Flavour Response              &  $-0.28 \pm 0.03$ & $+0.011$ &  $-0.008$ &  $+0.03 \pm 0.02$ & 0.14  & $-1.00$ \\   \hline       
   Close-by jets                 &  $-0.22 \pm 0.04$ & $+0.005$ &  $+0.002$ &  $+0.25 \pm 0.03$ & 0.01  & $-1.00$ \\   \hline      
    $b$-Jet energy scale         &  $+0.06 \pm 0.03$ & $+0.000$ &  $+0.010$ &  $+0.68 \pm 0.02$ & 0.34  & $+1.00$ \\ \hline        
   Total (without bJES)          &  0.58 $\pm$ 0.11 & 0.018   & 0.009  & 0.75 $\pm$ 0.08 & 0.41 & $-0.23$   \\ \hline
\end{tabular}
\end{center}
\caption{The individual components of the JES uncertainty according to
  Ref.~\protect\cite{CERN-PH-EP-2013-222}, together with the corresponding
  uncertainties on \mtlj, \JSF, \bJSF, \mtdl, and \mtcb.
 Some components listed are calculated as the sum in quadrature of several
 sub-components. The corresponding measurement correlations per group described
 in Sect.~\protect\ref{sec:resultcomb} are also reported.
\label{tab:jesresults}}
\end{table*}
\clearpage


\bibliographystyle{bibtex/bst/atlasBibStyleWithTitle}
\bibliography{TOPQ_2013_02,bibtex/bib/atlas-paper}

%

\newpage
\begin{flushleft}
{\Large The ATLAS Collaboration}

\bigskip

G.~Aad$^{\rm 85}$,
B.~Abbott$^{\rm 113}$,
J.~Abdallah$^{\rm 152}$,
O.~Abdinov$^{\rm 11}$,
R.~Aben$^{\rm 107}$,
M.~Abolins$^{\rm 90}$,
O.S.~AbouZeid$^{\rm 159}$,
H.~Abramowicz$^{\rm 154}$,
H.~Abreu$^{\rm 153}$,
R.~Abreu$^{\rm 30}$
% ...
\bigskip
\\
% ...
$^{11}$ Institute of Physics, Azerbaijan Academy of Sciences, Baku, Azerbaijan\\
% ...
$^{30}$ CERN, Geneva, Switzerland\\
% ...
$^{85}$ CPPM, Aix-Marseille Universit{\'e} and CNRS/IN2P3, Marseille, France\\
% ...
$^{90}$ Department of Physics and Astronomy, Michigan State University, East Lansing MI, United States of America\\
% ...
$^{107}$ Nikhef National Institute for Subatomic Physics and University of Amsterdam, Amsterdam, Netherlands\\
% ...
$^{113}$ Homer L. Dodge Department of Physics and Astronomy, University of Oklahoma, Norman OK, United States of America\\
% ...
$^{152}$ Institute of Physics, Academia Sinica, Taipei, Taiwan\\
$^{153}$ Department of Physics, Technion: Israel Institute of Technology, Haifa, Israel\\
$^{154}$ Raymond and Beverly Sackler School of Physics and Astronomy, Tel Aviv University, Tel Aviv, Israel\\
% ...
$^{159}$ Department of Physics, University of Toronto, Toronto ON, Canada\\
% ...
\end{flushleft}


\end{document}